\documentclass[11pt]{article}

\newcommand{\vs}[1]{\vspace{#1 mm}}

\newcommand{\mathbb}[1]{{\bf{#1}}}
\newcommand{\text}[1]{{{#1}}}

\newcommand{\notag}{\nonumber}
\newcommand{\func}[1]{{#1}}

\newcommand{\frak}[1]{{\bf{#1}}}

\begin{document}

\begin{titlepage}
\begin{center}

\hfill  {LPTENS-01/04} \\
\hfill  {HUTP-01/A002} \\
\hfill  {\tt hep-th/0101219} \\
        [.3in]

{\large\bf
NONASSOCIATIVE STAR PRODUCT DEFORMATIONS FOR $D$--BRANE WORLDVOLUMES IN
CURVED BACKGROUNDS
}\\[.3in]

{\bf Lorenzo Cornalba}\footnote{E-mail: {\tt cornalba@lpt.ens.fr}}
$\;$ and $\;$
{\bf Ricardo Schiappa}\footnote{E-mail: {\tt ricardo@lorentz.harvard.edu}}\\

\vs{4}
{\it 1 -- Laboratoire de Physique Th\'eorique}\\
{\it \'Ecole Normale Sup\'erieure}\\
{\it F-75231 Paris Cedex 05, FRANCE}\\
\vs{2}
{\it 2 -- Department of Physics}\\
{\it Harvard University}\\
{\it Cambridge, MA 02138, U.S.A.}\\

\end{center}

\vs{1}
\centerline{{\bf{Abstract}}}
\vs{1}

\small{
\noindent
We investigate the deformation of $D$--brane world--volumes in curved
backgrounds. We calculate the leading corrections to the boundary conformal
field theory involving the background fields, and in particular we study
the correlation functions of the resulting system. This allows us to obtain
the world--volume deformation, identifying the open string metric and the
noncommutative deformation parameter. The picture that unfolds is the
following: when the gauge invariant combination $\omega = B + F$ is
constant one obtains the standard Moyal deformation of the brane
world--volume. Similarly, when $d\omega = 0$ one obtains the noncommutative
Kontsevich deformation, physically corresponding to a curved brane in a
flat background.  When the background is curved, $H = d\omega \not = 0$, we
find that the relevant algebraic structure is still based on the Kontsevich
expansion, which now defines a nonassociative star product with an 
$A_{\infty}$ homotopy associative algebraic structure. We then
recover, within this formalism, some known results of Matrix theory in
curved backgrounds. In particular, we show how the effective action
obtained in this framework describes, as expected, the dielectric effect of
$D$--branes. The polarized branes are interpreted as a soliton, associated
to the condensation of the brane gauge field.
}

\vs{2}

\noindent
\hfill January 2001

\end{titlepage}

\renewcommand{\thefootnote}{\arabic{footnote}} \setcounter{footnote}{0}

\newpage \begin{tableofcontents}
\end{tableofcontents}
\pagebreak

\section{Introduction and Summary}

Noncommutative quantum field theoretic limits of string theory have
received considerable attention in the recent literature, and have been
studied in a variety of papers (see, \textit{e.g.}, \cite{CDS,
Cheung-Krogh, Schomerus, Cornalba-Schiappa, Seiberg-Witten, Cornalba-2} and
references therein). The attention is focused on a specific scaling limit,
where the effects of large magnetic backgrounds are translated into Moyal
noncommutative deformations of the $D$--brane world--volume algebra of
functions. The open string physics is therefore captured within a quantum
field theory (which is renormalizable, despite appearances \cite{MRS,
MST}). A common point to most previous investigations is that the
background (sigma model) fields are taken to be constant and that, as a
consequence, the target space is flat. One may then ask the natural
question of what happens if the background is curved, \textit{i.e.}, if the
background fields are no longer constant?  This question received some
attention in a couple of recent papers \cite{Anazawa, ARS-1, ARS-2,
Ho-Yeh}, but there is no general answer to it (other papers of interest
with some relation to this subject are, \textit{e.g.}, \cite{Chu-Ho-Kao,
Ydri, Sahakian, Dasgupta-Yin}). Our goal in this work is to address this
problem in the context of a simple model with weakly curved backgrounds,
which can be on one side connected to the known flat background framework,
and on the other hand can be related to formal results of brane physics in
WZW models, which can be analyzed exactly with conformal field theory
techniques \cite{ARS-1, ARS-2, BSD}.

More concretely, the aim of this paper is to understand how the presence of
a non--trivial background field affects the world--volume deformation of a
$D$--brane. It is known that, in the presence of a constant background
$B$--field, the physics can be exactly described either by a sigma model
approach \cite{AFM, Fradkin-Tseytlin, BCZ, Dorn-Otto, ACNY, CLNY, Tseytlin,
Laidlaw}, or alternatively, by translating the background $B$--field into a
noncommutative Moyal deformation of the brane world--volume algebra of
functions \cite{Schomerus, Seiberg-Witten}. The constant field situation
represents a particular choice of background and one can ask what happens
in more complicated situations. One thing to keep in mind is that (as for
the Born--Infeld action \cite{Leigh, Polchinski, Witten-2}) the gauge
covariant combination to consider is not $B$ alone, but $B+F$, which we
shall denote by $\omega \equiv B+F$ in the following. One may then consider
three cases of increasing complexity: the case of constant $\omega$, the
case where $d\omega=0$ but $\omega$ is \textit{not} constant, and the most
general case where $d\omega \not = 0$ and we have NS--NS three form flux
(as $d\omega = dB + dF = H$) and a curved background.

The analysis leads to the following complete picture. The first case,
corresponding to constant $\omega$, has been extensively studied in the
literature where one obtains a noncommutative Moyal deformation of the
brane world--volume \cite{Schomerus, Cornalba-Schiappa, Seiberg-Witten,
Cornalba-2}. The physics is by now very well understood, corresponding to a
flat brane embedded in a flat background space. The second case, when
$\omega$ is not constant but $d\omega=0$, has also been studied in the
literature, though to a much less extent. This gives the so--called
Cattaneo--Felder model of \cite{Cattaneo-Felder}. One therefore obtains the
natural extension of the Moyal deformation to the case of varying
symplectic form, corresponding to the noncommutative Kontsevich star
product deformation of the brane world--volume algebra of functions
\cite{Kontsevich}. This situation corresponds to the embedding of a curved
brane in a flat background space. These configurations have also been
studied from the point of view of BPS membranes in Matrix theory, where the
varying $F$--field physically corresponds to a varying density of
zero--branes over a curved membrane \cite{Cornalba-Taylor, Cornalba-1}.
Finally, the general case where $d\omega \not = 0$ is the main subject of
this paper. One no longer has a symplectic form and apparently no obvious
definition of a star product ---which usually comes from a given Poisson
structure on the world--volume of the $D$--brane.  In this general
situation, we will find that the world--volume algebra of functions is
deformed to an algebra which is not only noncommutative, but also
nonassociative. One interesting point we shall uncover is that this
nonassociative star product can still be defined using Kontsevich's formula
\cite{Kontsevich}. Therefore, the nonassociativity can be traced, thanks to
Kontsevich's formality formulae, to the Schouten--Nijenhuis bracket of
$\omega^{-1}$ with itself, which is proportional to the NS--NS
field strength $d\omega= H$ \cite{Kontsevich, Cattaneo-Felder}. These
nonassociative algebras have the structure of an $A_{\infty}$ homotopy 
associative algebra (see, \textit{e.g.}, \cite{Stasheff, Lada-Stasheff, 
Lada-Markl}) which have previously received some attention in the string 
field theory literature since they are the natural algebras that appear in 
general open--closed string field theories \cite{Zwiebach,GAB}.

Our approach in this paper will rely on a perturbative calculation of
$n$--point functions on the disk, using the background field method applied
to open string theory \cite{AFM, Fradkin-Tseytlin, ACNY, CLNY}. The
background fields are expanded in Taylor series, and the derivative terms
that appear are treated as new interactions, which we treat in a
perturbative expansion. This allows us to obtain the open string
parameters, metric $G$ and deformation $\theta$, generalizing the results
in \cite{CLNY, Schomerus, Seiberg-Witten}. It also allows us to identify
the star product deformations, as described in the previous paragraph. We
begin, in section 2, by describing the specific closed string backgrounds
which we shall consider in this paper. These will be the class of
parallelizable manifolds, exact background solutions for closed string
theory \cite{BCZ}. Then, in section 3, we shall describe in detail the
perturbation theory on the disk for open strings in these curved
backgrounds, \textit{i.e.}, we will study the new interaction vertices due
to the curvature terms. In particular, we present the general methods that
we then use in section 4 for the calculation of $n$--point functions on the
disk, with particular emphasis on the conformal properties of these disk
correlators. These correlators also yield the open string parameters and
the nonassociative Kontsevich star product. Section 5 includes a brief
resume of the different situations and the different world--volume
deformations and star products, which can be read directly by the reader
who wishes to skip the calculations in the preceding sections.  It also
describes in some detail the concept of a nonassociative star product
deformation, which could be a topic of great interest for future research.

Most of the previous treatment is done in a particular $\alpha' \to 0$
scaling limit \cite{Seiberg-Witten}, where the closed string metric, $g$,
scales to zero. In section 6 we move away from this limit and compute
corrections to the previous results which explicitly depend on the closed
string metric. These calculations yield the formulas relating open and
closed string parameters. It is interesting to observe that the final
answer is a simple generalization of the flat background results of
\cite{CLNY, Schomerus, Seiberg-Witten}. In section 7 we make contact with
previous results and in particular we describe, within our formalism, the
dielectric effect of $D$--branes \cite{Myers} in these curved backgrounds.
Indeed, these solutions describing polarization of lower dimensional
branes, obtained first in \cite{Myers} and then further studied in
different situations involving $D$--branes and fundamental strings in R--R
or NS--NS backgrounds by, \textit{e.g.}, \cite{CMT, Schiappa, MTR, JMR,
DTV, Kluson, Lozano}, is now reinterpreted, dually, as an instability of
the space filling brane, which condenses to a lower dimensional brane.
This is accomplished by first studying the relation between the partition
function ---the correlators we computed in the earlier sections--- and the
effective action. Once this connection is made (using boundary string field
theory arguments), we obtain the usual matrix action in the presence of an
$H$--field, and we can then use the previous results on the subject.

Finally, we discuss in the concluding sections how further studies of
these nonassociative geometries could lead to a proper definition of Matrix
theory \cite{Witten-1, BFSS, DHN, Sen, Seiberg} in a general curved
background. These nonassociative geometries could provide the proper
framework to generalize the arguments in \cite{Douglas-Talk, Douglas-1,
DKO} and the weak field calculations of \cite{Taylor-Raamsdonk-1, DNP} in
order to build the Matrix theory action in a general curved target space.

\section{Open Strings in Parallelizable Backgrounds}

The physics of a string propagating in a curved background is conveniently
described in terms of a nonlinear sigma model. In the presence of a
background metric $g_{ab}\left( x\right)$ and NS--NS 2--form field $B_{ab} 
\left( x\right)$ the action which governs the motion of the string is
given by \cite{AFM, Fradkin-Tseytlin, BCZ, ACNY, CLNY},

\begin{equation}
S=\frac{1}{4\pi \alpha ^{\prime }}\int_{\Sigma }g_{ab}\left( X\right)
dX^{a}\wedge \ast dX^{b}+\frac{i}{4\pi \alpha ^{\prime }}\int_{\Sigma
}B_{ab}\left( X\right) dX^{a}\wedge dX^{b},  \label{open-string}
\end{equation}

\noindent
where $\Sigma$ is the string world--volume. Moreover, when 
considering open strings one can include boundary interactions on $\partial 
\Sigma$. In the sequel, we will mainly focus on the coupling to the $U\left( 
1\right)$ gauge field $A_{a}\left( x\right)$, given by 

$$
S_{B}=i\oint_{\partial \Sigma }A_{a}\left( X\right) dX^{a}.
$$

\noindent
In this paper we will consider only the physics at weak string coupling, and 
we will consequently assume $\Sigma $ to have the topology of a
disk. Other background fields (such as the dilaton) will not play a 
role in our subsequent analysis. We shall mainly address maximal 
branes, though our results are completely general. Also, from now on, we will 
work in units such that

$$
2\pi \alpha ^{\prime }=1.
$$

The action (\ref{open-string}) is written in a generic coordinate system 
$x^{a}$ in spacetime. On the other hand, in order to use (\ref{open-string}) 
to compute correlators in perturbation theory, it is natural to follow the
standard techniques of the background field method and use coordinates 
$x^{a}$ which are \textit{Riemann normal coordinates at the origin} 
---\textit{i.e.} defined using geodesic paths in target space which start 
at $x^{a}=0$ \cite{AFM, BCZ}. We recall that the main advantage of this 
choice is that the Taylor series expansion of any tensor around $x^{a}=0$ 
is explicitly given in terms of covariant tensors evaluated at the origin. 
In particular one has, up to quadratic order in the coordinates, 

\begin{equation}
g_{ab}\left( x\right) =g_{ab}-\frac{1}{3}R_{acbd}x^{c}x^{d}+\cdots .
\label{metricEx}
\end{equation}

\noindent
Let us now consider the expansion of the NS--NS $2$--form field, by first
recalling that we have some gauge freedom in the definition of $B_{ab}\left(
x\right)$. In fact, the transformations $B\rightarrow B+d\Lambda$, 
$A\rightarrow A-\Lambda$ leave the total action $S+S_{B}$ invariant, and we
can use this freedom to impose the following (radial) gauge\footnote{Given 
a generic field $B_{ab}\left( x\right) $, we can consider the gauge
transformation parameter $\Lambda _{a}\left( x\right) $ given by $\Lambda
_{a}\left( x\right) =x^{b}\int_{0}^{1}sB_{ab}\left( sx\right) ds$. It is
then a simple computation to see that the combination $\partial _{a}\Lambda
_{b}-\partial _{b}\Lambda _{a}$ equals $-B_{ab}\left( x\right)
+x^{c}\int_{0}^{1}s^{2}H_{abc}\left( sx\right) ds$.} 

$$
x^{a}B_{ab}\left( x\right) =x^{a}B_{ab}\left( 0\right) .
$$

\noindent
One can explicitly solve the above equation in terms of the NS--NS 
three--form field strength

$$
H=dB,
$$

\noindent
and obtain 

$$
B_{ab}\left( x\right) =B_{ab}+x^{c}\int_{0}^{1}s^{2}H_{abc}\left( sx\right)
ds.
$$

\noindent
Therefore, the normal coordinate expansion for the field $B_{ab}$ is
explicitly given by

\begin{equation}
B_{ab}\left( x\right) =B_{ab}+\frac{1}{3}H_{abc}x^{c}+\frac{1}{4}\nabla
_{d}H_{abc}x^{c}x^{d}+\cdots .
\label{BEx}
\end{equation}

Using the expressions (\ref{metricEx}) and (\ref{BEx}), one can expand 
(\ref{open-string}) about the classical constant background $\partial X^{a}=0$ and 
obtain

\begin{equation}
S=S_{0}+S_{1}+\cdots ,
\label{exp}
\end{equation}

\noindent
where $S_{n}$ contains $n+2$ powers of the coordinate fields $X^{a}$ and
where, in particular, 

\begin{eqnarray}
S_{0} &=&\frac{1}{2}g_{ab}\int_{\Sigma }dX^{a}\wedge \ast dX^{b}+\frac{i}{2}
B_{ab}\int_{\Sigma }dX^{a}\wedge dX^{b} , \notag \\
S_{1} &=&\frac{i}{6}H_{abc}\int_{\Sigma }X^{a}dX^{b}\wedge dX^{c}.
\label{a1}
\end{eqnarray}

In this paper, we will be primarily interested in the effects of the term 
$S_{1}$, which describes a small curved deviation from the flat closed 
string background. Let us elaborate more on this point. To leading order 
in $\alpha^{\prime}$, the beta function equations which describe consistent
closed string backgrounds read \cite{AFM, BCZ}:

\begin{equation}
R_{ab}=\frac{1}{4}H_{acd}{H_{b}}^{cd}\ ,\ \ \ \ \ \ \ \ \ \ \ \ \ \ \ \
\nabla ^{a}H_{abc}=0.
\label{e200}
\end{equation}

\noindent
If we work to first order in $H$, one may then neglect the presence of
curvature coming from the metric and only consider the effects of $H$
coming from (\ref{a1}). We can actually make these arguments more systematic
if we consider a general class of conjectured solutions to the beta 
function equations, called parallelizable manifolds \cite{BCZ}. These 
configurations are characterized by the following properties. First of all, 
the tensor $H_{abc}$ is covariantly constant,

$$
\nabla _{a}H_{bcd}=0.
$$

\noindent
Moreover, if we consider the generalized connection $\Gamma +\frac{1}{2}H$,
then the corresponding curvature tensor,

$$
\mathcal{R}_{abcd}=R_{abcd}+\frac{1}{2}\nabla_{a}H{}_{bcd}-\frac{1}{2}
\nabla_{b}H{}_{acd}+\frac{1}{4}H_{ade}{H_{bc}}^{e}-\frac{1}{4} 
H_{ace}{H_{bd}}^{e}\ ,
$$

\noindent
must vanish. Using the fact that $R_{a[bcd]}=0$, one can easily show that
the field $H_{abc}$ must satisfy a Jacobi identity, in the sense that 

$$
H_{abe}{H_{cd}}^{e}+\mathrm{cyclic}_{abc}=0.
$$

\noindent
These facts then imply 

$$
R_{abcd}=\frac{1}{4}H_{abe}{H_{cd}}^{e}\ ,
$$

\noindent
and therefore (\ref{e200}). Moreover, at a more fundamental level, it was
explicitly shown that when the target is parallelizable, the string sigma
model is ultra--violet finite to two loops, with vanishing beta 
functions \cite{BCZ}. It was moreover suggested that this holds true to 
higher orders for the superstring, and one thus has a consistent solution of 
closed string theory \cite{BCZ}.

In the parallelizable situation the expansion (\ref{exp}) drastically
simplifies. In the sequel we shall only need the explicit forms of $S_{0}$
and $S_{1}$ given above. On the other hand, in order to extend the results of 
this paper to higher order in $H$, one needs the expressions of $S_{n}$ for 
$n\geq 2$. We include, for completeness, the first of these terms explicitly
given by:

$$
S_{2}=-\frac{1}{24}H_{abe}{H_{cd}}^{e}\int_{\Sigma }X^{a}X^{c}dX^{b} 
\wedge \ast dX^{d}.
$$

\section{Perturbation Theory}

In the last section we have reviewed the general form of the sigma model
action which describes open string dynamics in curved backgrounds. From now
on we shall only consider backgrounds which are weakly curved. More
precisely we will work, for the rest of the paper, to \textit{leading} order
in the background field $H$, and consequently we shall focus our 
analysis on the action $S_{0} + S_{1} + S_{B}$. If we denote with $F=dA$ the 
$U\left( 1\right) $ field strength, and with $\omega$ the symplectic 
structure

\begin{equation}
\omega_{ab}\left( x\right) = B_{ab}+F_{ab}\left( x\right) ,
\label{omega}
\end{equation}

\noindent
then the relevant action is given by 

\begin{equation}
\frac{1}{2}g_{ab} \int_{\Sigma } dX^{a} \wedge \ast dX^{b} + i \int_{\Sigma}
\omega + \frac{i}{6} H_{abc} \int_{\Sigma} X^{a} dX^{b} \wedge dX^{c}.
\label{mainAct}
\end{equation}

Before we start the detailed discussion of the perturbation theory for the
action (\ref{mainAct}), and in order to set the stage and motivate the
subsequent results, let us begin by recalling some known facts which are 
valid in the flat space limit of $H_{abc}=0$. On one side, the conventional 
approach to open string physics starts by considering the simple free action 
$\frac{1}{2} g_{ab} \int_{\Sigma} dX^{a} \wedge \ast dX^{b}$, or even the full 
free action $S_{0}$. One then analyzes the physics of boundary interactions by
considering the coupling $S_{B}$ to the $U(1)$ gauge field, $A$, and one
treats (following, for example, the approaches in \cite{Fradkin-Tseytlin, 
ACNY, CLNY}) the interactions perturbatively in $F=dA$. In this scheme the 
basic interaction vertex with $n$ external legs involves $n-2$ derivatives of 
$F$, and the perturbation theory quickly becomes unmanageable as soon as one 
considers rapidly varying gauge fields. It was noted, on the other hand, in 
\cite{Cattaneo-Felder} that, if one considers the simple topological action $i 
\int_{\Sigma} \omega$ (that is, one looks at (\ref{mainAct}) in the limit 
$g_{ab}$, $H_{abc} \to 0$), then the resulting path integral drastically 
simplifies. In fact, if one considers the $n$--point function of $n$ generic 
functions $f_{1}\left( x\right)$, \ldots, $f_{n}\left( x\right)$, placed 
cyclically on the boundary $\partial \Sigma$ of the string world--volume, one 
obtains the simple result (independently of the moduli of the insertion 
points) \cite{Cattaneo-Felder,Seiberg-Witten}: 

\begin{equation}
\langle f_{1} \cdots f_{n} \rangle = \int V\left( \omega \right) dx\, 
\left( f_{1} \star \cdots \star f_{n}\right) .
\label{nptfct}
\end{equation}

\noindent
In the above, $\star$ is the \textit{associative} Kontsevich star 
product\footnote{The terms hidden behind the dots $\cdots$ in (\ref{Kont}) are 
given by explicit diagrammatic expressions, as explained in \cite{Kontsevich}, 
valid for any bi--vector field $\alpha^{ab} \left( x\right)$ in terms of the 
functions $f$, $g$, the tensor $\alpha^{ab}$ and their derivatives. If 
$\alpha^{-1}$ is closed, then the corresponding product is associative.} 
with respect to the Poisson structure $\alpha = \omega^{-1}$,

\begin{equation}
f\star g=f\cdot g+\frac{i}{2}\alpha ^{ab}\partial _{a}f\,\partial
_{b}g+\cdots ,  \label{Kont}
\end{equation}

\noindent
and $V\left( \omega \right) = \sqrt{\det \omega} \left( 1 + 
\cdots \right)$ is a volume form\footnote{For more details on $V\left( 
\omega \right) $ we refer the reader to \cite{Cornalba-3}.} such that 
$\int V\left( \omega \right)\, dx$ acts as a trace for the product 
$\star$. The basic point we would like to stress is that the product 
(\ref{Kont}) contains derivatives of $\alpha$ (and therefore of $F$) 
to all orders, and is therefore valid for arbitrary gauge field 
configurations. This means that the perturbation theory in $A_{a}$ becomes 
tractable to all orders when $g_{ab} \to 0$, and is conveniently 
described in terms of the algebraic operation~$\star$.

We shall see in this paper that, when one introduces the perturbation 
$S_{1}$ but still considers the limit $g_{ab}\rightarrow 0$, then one can still
re--sum the perturbation theory to all orders in $A_{a}$. We will see that the
relevant algebraic structure is still given by a Kontsevich product of the
general form (\ref{Kont}), but now with $\omega$ replaced in a natural
way by the gauge invariant combination:

$$
\widetilde{\omega}_{ab}\left( x\right) = \omega_{ab}\left( x\right) + 
\frac{1}{3}H_{abc}x^{c} = B_{ab}\left( x\right) + F_{ab}\left( 
x\right) ,
$$

\noindent
and with $\alpha$ replaced by $\widetilde{\alpha} =
\widetilde{\omega}^{-1}$.  In order to clearly distinguish the two cases,
we shall denote this second product (relative to $\widetilde{\alpha}$) with
$\bullet$, given by the usual Kontsevich expansion,

$$
f \bullet g = f\cdot g + \frac{i}{2}\widetilde{\alpha}^{ab}\partial_{a}f
\,\partial_{b}g + \cdots .
$$

\noindent
The two--form $\widetilde{\omega}$ is not closed and correspondingly the
product $\bullet $ is now \textit{nonassociative}. We will discuss later
how the nonassociativity is controlled by the field strength $H =
d\widetilde{\omega}$. The $n$--point functions are again given by an
equation similar to (\ref{nptfct}), with $\star$ replaced by $\bullet$. On
the other hand, expressions like $f_{1}\bullet \cdots \bullet f_{n}$ are
ambiguous, due to the nonassociativity of the product, and one needs to
insert parenthesis to precisely define their meaning. This can be done in
various ways, and this fact is reflected in the dependence of $n$--point
functions on the $n-3$ \textit{conformal} moduli of the insertion points on
the boundary $\partial \Sigma$. The $n$--point functions will then be
interpolations, parameterized by $n-3$ moduli, between the various possible
positions of the parenthesis in the expression $f_{1}\bullet \cdots \bullet
f_{n}$.

From now until section 4.5 we will concentrate on the simplest case of 
$F=0$ or

$$
\omega _{ab}\left( x\right) =B_{ab}.
$$

\noindent
We thus neglect the boundary interaction $S_{B}$ and concentrate on the
action $S_{0} + S_{1}$. The generalization to the case (\ref{omega}) will be
comparatively simple (as for the $d\widetilde\omega = 0$ case)
and is left to section 4.5, which also summarizes
the results in the general context. We now turn to a systematic discussion
of the perturbation theory for the action $S_{0} + S_{1}$.

\subsection{The Free Theory}

Let us first recall some facts about the unperturbed action $S_{0}$. Since 
$S_{0}$ is invariant under translations $X^{a}\rightarrow X^{a}+c^{a}$, the
field $X^{a}$ can be split into a constant zero mode $x^{a}$ and a
fluctuating quantum field $\zeta^{a}$,

\begin{equation}
X^{a}=x^{a}+\zeta ^{a}.
\label{split}
\end{equation}

\noindent
Path integrals with the free action $S_{0}$ are then explicitly given by a
path integral over the quantum field $\zeta^{a}$ and an ordinary integral
over the zero--mode $x^{a}$ as \cite{Fradkin-Tseytlin}:

$$
\int [dX]\ e^{-S_{0}\left( X\right) }\rightarrow \int dx \int 
[d\zeta]\ e^{-S_{0}\left( \zeta \right) }.
$$

The integral in $[d\zeta]$ is gaussian and is determined once one obtains 
the two--point function for the fluctuating field $\zeta$. From now on, 
and unless otherwise specified, we will parameterize the disk $\Sigma$ with 
the complex upper--half plane $\mathbb{H}^{+}$. As discussed in \cite{ACNY, 
Schomerus, Seiberg-Witten}, the two--point function can be more 
conveniently written if one introduces the open string metric $G_{ab}$ and 
noncommutativity tensor $\theta^{ab}$ as given by

$$
\frac{1}{G}+\theta =\frac{1}{g+B}.
$$

\noindent
It then has the general form,

\begin{equation}
\langle \zeta^{a} \left( z\right) \zeta^{b} \left( w\right) \rangle = 
\frac{i}{\pi} \theta^{ab} \mathcal{A}\left( z,w\right) - 
\frac{1}{\pi} G^{ab} \mathcal{B}\left( z,w\right) + 
\frac{1}{2\pi} g^{ab} \mathcal{C}\left( z,w\right) ,
\label{prop}
\end{equation}

\noindent
where 

\begin{eqnarray*}
\mathcal{A}\left( z,w\right) &=& \frac{1}{2i} \ln \left( 
\frac{\overline{w}-z} {\overline{z}-w} \right) , \\
\mathcal{B}\left( z,w\right) &=& \ln \left| z-\overline{w} \right| , \\
\mathcal{C}\left( z,w\right) &=& \ln \left| \frac{z-w}{z-\overline{w}} 
\right| .
\end{eqnarray*}

\noindent
In the sequel, we shall only need to consider the propagator (\ref{prop})
when one point (say $w$) is placed at the boundary $\partial \Sigma$ of the
string world--sheet. In this case $w=\overline{w}$ and $\mathcal{C}\left(
z,w\right) = 0$. Also, in the case $w=\overline{w}$, the coefficients 
$\mathcal{A}\left( z,w\right)$ and $\mathcal{B}\left( z,w\right)$ have a
simple geometrical interpretation. $\mathcal{A}$ measures the angle 
between the line $z$--$w$ and the vertical line passing through $w$, and 
$\mathcal{B}$ gives the logarithm of the distance between $z$ and $w$.

We now consider the limit $g_{ab}\rightarrow 0$. In this limit, the
effective open string metric $G_{ab}$ becomes large and therefore the term
in (\ref{prop}) proportional to $G^{ab}$ becomes irrelevant. Also, one 
has in this limit, that $\theta = B^{-1} = \alpha(x)$. In this case the 
propagator (\ref{prop}) reduces to

$$
\langle \zeta^{a}\left( z\right) \zeta^{b}\left( w\right) \rangle 
= \frac{i}{\pi} \theta^{ab} \mathcal{A}\left( z,w\right) ,
$$

\noindent
and the computation of path integrals becomes simple. As we discussed in the
previous subsection, if one considers $n$ functions $f_{1}$, \ldots, 
$f_{n}$, positioned at ordered points $\tau _{1} < \cdots < \tau _{n}$ on the 
boundary $\partial \Sigma$ of the string world--sheet, then the path integral

\begin{equation}
\int [dX]\ e^{-S_{0} \left( X\right) } \, f_{1} \left( X\left( \tau 
_{1} \right) \right) \cdots f_{n} \left( X\left( \tau _{n} \right) \right)
\label{free}
\end{equation}

\noindent
can be evaluated \cite{Schomerus, Seiberg-Witten} with the result,

$$
\int V\left( B\right) \,dx\,\left( f_{1}\star \cdots \star f_{n}\right) .
$$

\noindent
Since $\omega \left( x\right) = B$ is constant, the product $\star$ is the
usual Moyal star product and $V\left( B\right) = \sqrt{ \det B }$.

A word on notation. From now on we will omit the explicit reference to the
volume form in the integrals. We shall therefore use the following 
short--hand notation:

$$
\int V\left( \omega \right) \,dx\,\cdots \rightarrow \int \cdots .
$$

\subsection{The Interaction}

Let us now consider the effects of the perturbation $S_{1}$. Corresponding
to the split (\ref{split}), the effect of $S_{1}$ is to introduce two bulk
graphs:

\begin{eqnarray}
\mathcal{V} &=& -\frac{i}{6} H_{abc}x^{c} \int_{\Sigma} d\zeta^{a} \wedge
d\zeta^{b}\ , \label{V2} \\
\mathcal{W} &=& -\frac{i}{6} H_{abc} \int_{\Sigma} \zeta^{a} d\zeta^{b} 
\wedge d\zeta^{c}. \label{V3}
\end{eqnarray}

\noindent
We will then consider the following path integral

\begin{eqnarray}
&& \int [dX]\ e^{-S_{0}\left( X\right) - S_{1}\left( X\right) }\,f_{1} 
\left( X \left( \tau _{1} \right) \right) \cdots f_{n} \left( X \left( 
\tau_{n} \right) \right) \simeq \nonumber \\
&& \int [dX]\ e^{-S_{0}\left( X\right)}\, \left[ 1 + \mathcal{V} + 
\mathcal{W} \right] \,\, f_{1} \left( X \left( \tau_{1} \right) \right) 
\cdots f_{n}\left( X \left( \tau_{n} \right) \right) .
\label{pathInt}
\end{eqnarray}

In order to analyze the effects of $\mathcal{V}$ and $\mathcal{W}$, let us
first introduce some notation and discuss some useful simple results. 
Consider a generic point $z\in \mathbb{H}^{+}$, and consider the path 
integral:

$$
\int [dX]\ e^{-S_{0}\left( X\right) }\,\zeta ^{a}\left( z\right)
\,\,f_{1}\left( X\left( \tau _{1}\right) \right) \cdots f_{n}\left( X\left(
\tau _{n}\right) \right).
$$

\noindent
If we introduce the short--hand notation,

$$
\mathcal{A}\left( z,\tau _{i}\right) =\mathcal{A}_{i} ,
$$

\noindent
for the angle between the line $z$--$\tau_{i}$ and the vertical through 
$\tau_{i}$, then the result of the above path integral is simply given 
by

$$
\sum_{i=1}^{n}\frac{i}{\pi }\mathcal{A}_{i}\,\theta ^{a\widetilde{a}}\int
f_{1}\star \cdots \star \partial _{\widetilde{a}}f_{i}\star \cdots \star
f_{n}.
$$

\noindent
The above result is easy to understand once one considers the expansion of 
the functions $f_{i}\left( X\right) =f_{i}\left( x+\zeta \right)$ in Taylor
series in powers of $\zeta$. The contraction of the field $\zeta^{a} \left(
z\right)$ with a field $\zeta^{\widetilde{a}} \left( \tau _{i}\right)$
coming from the Taylor expansion of the function $f_{i}$ gives a factor of 
$\frac{i}{\pi} \mathcal{A}_{i}\,\theta ^{a\widetilde{a}}$. We are then left
with a path integral of the form (\ref{free}), where the function $f_{i}$
has been replaced with its derivative $\partial_{\widetilde{a}} f_{i}$. More
generally, when a free field $\zeta^{a} \left( z \right)$ is contracted with
one of the boundary functions it acts as a differentiation:

\begin{equation}
\frac{i}{\pi} \mathcal{A\,} \theta^{a\widetilde{a}} 
\partial_{\widetilde{a}} \ .
\label{contraction}
\end{equation}

With this result, we can now consider the effects of the perturbation
vertices $\mathcal{V}$ and $\mathcal{W}$ in the path integral (\ref
{pathInt}). Let us start with the analysis of $\mathcal{V}$. Choose 
any two indices $i<j$ and consider the term where the two $\zeta$'s in 
$\mathcal{V}$ differentiate the two functions $f_{i}$ and $f_{j}$ (in the 
sense just described above). If $\zeta^{a}$ differentiates $f_{i}$ and 
$\zeta^{b}$ differentiates $f_{j}$ one then gets

$$
\frac{i}{6\pi ^{2}}H_{abc}\theta ^{a\widetilde{a}}\theta ^{b\widetilde{b}}
\left( \int_{\Sigma }d\mathcal{A}_{i}\wedge d\mathcal{A}_{j}\right) \int
x^{c}\left( \cdots \star \partial _{\widetilde{a}}f_{i}\star \cdots \star
\partial _{\widetilde{b}}f_{j}\star \cdots \right) .
$$

\noindent
The integral over $\Sigma $ can be evaluated by noting that the upper--half
plane $\mathbb{H}^{+}$ corresponds to the simplex $-\frac{\pi}{2} < 
\mathcal{A}_{i} < \mathcal{A}_{j} < \frac{\pi}{2}$ in the 
$\mathcal{A}_{i}$--$\mathcal{A}_{j}$ plane. Therefore the integral 
$\int_{\Sigma} d\mathcal{A}_{i} \wedge d\mathcal{A}_{j}$ is equal to 
$\frac{1}{2}\pi^{2}$. Moreover, if we instead let $\zeta^{a}$ differentiate 
$f_{j}$ and $\zeta^{b}$ differentiate $f_{i}$ we obtain, using the 
antisymmetry of $H_{abc}$, the same result as above. Summing the two 
contributions, and summing over all possible pairs $i<j$, one then 
obtains

$$
\sum_{i<j}V_{ij} ,
$$

\noindent
where, for $i<j$, we have defined 

$$
V_{ij}=\frac{i}{6}H_{abc} \theta^{a\widetilde{a}} \theta^{b\widetilde{b}}
\int x^{c}\star \left( f_{1}\star \cdots \star \partial_{\widetilde{a}}
f_{i}\star \cdots \star \partial_{\widetilde{b}} f_{j}\star \cdots \star
f_{n}\right) .
$$

\noindent
In the above equation we have used the fact that (for the Moyal product) 
$\int f \cdot g=\int f\star g$, in order to rewrite everything in terms 
of $\star$ products, including the multiplication by the coordinate function 
$x^{c}$. To conclude the analysis of the effect of the two--vertex 
$\mathcal{V}$, one must also consider the term coming from the contraction of 
the two $\zeta$'s in $\mathcal{V}$ among themselves. This term will require 
some care, since we must regularize the contraction of two fields at 
coincident points. On the other hand, the general structure of the contribution 
can be obtained with little effort by recalling that the two indices $a$ and 
$b$ in (\ref{V2}) are contracted with the antisymmetric tensor $H_{abc}$. 
This implies that the contribution in question must have the form 

$$
V=\mathcal{N\,}H_{abc}\theta ^{bc}\int x^{a}\star \left( f_{1}\star \cdots
\star f_{n}\right) ,
$$

\noindent
where $\mathcal{N}$ is an unknown constant which will later be determined 
to be $1/3$.

We now move to the analysis of the contributions coming from the
three--graph $\mathcal{W}$. First, given three indices $i<j<k$, let us
define 

$$
W_{ijk}=-\frac{1}{12}H_{abc}\theta ^{a\widetilde{a}}\theta ^{b\widetilde{b}}
\theta ^{c\widetilde{c}}\int f_{1}\star \cdots \star \partial _{\widetilde{a}}
f_{i}\star \cdots \star \partial _{\widetilde{b}}f_{j}\star \cdots \star
\partial _{\widetilde{c}}f_{k}\star \cdots \star f_{n}.
$$

\noindent
It is then easy to check, using the general result (\ref{contraction}), that
the contribution from the three--vertex which comes from the contraction of
the fields $\zeta $'s in $\mathcal{W}$ with the functions $f_{i},f_{j},f_{k}$
is given by

$$
\sum_{i<j<k} S\left( \tau _{i},\tau _{j},\tau _{k}\right) W_{ijk},
$$

\noindent
where the function $S$ is 

\begin{eqnarray*}
S\left( \tau _{i},\tau _{j},\tau _{k}\right) &=&\frac{2}{\pi ^{3}}\left[
\int_{\Sigma }\mathcal{A}_{i}d\mathcal{A}_{j}\wedge d\mathcal{A}_{k}\pm 
\text{\textrm{permutation}}_{ijk}\right] \\
&=&\frac{4}{\pi ^{3}}\left[ \int_{\Sigma }\mathcal{A}_{i}d\mathcal{A}_{j}
\wedge d\mathcal{A}_{k}+\mathrm{cyclic}_{ijk}\right] .
\end{eqnarray*}

\noindent
Other combinations, which involve contractions of the $\zeta$'s 
amongst themselves, yield a vanishing contribution to the result of the 
three--vertex $\mathcal{W}$.

Let us analyze the function $S$ in more detail. As we saw, it depends on
three ordered points $\tau _{i}<\tau _{j}<\tau _{k}$ on the boundary 
$\partial \Sigma $. On the other hand, since it is written explicitly in
terms of integrals of angle functions $\mathcal{A}$, it is actually
invariant under translations $\tau \rightarrow \tau +c$ and scalings 
$\tau \rightarrow \lambda \tau $, \textit{i.e.}, under the subgroup of the
modular group $SL\left( 2,\mathbb{R}\right) $ which leaves invariant the
point at infinity. Therefore, by sending $\tau _{i}$ to $0$ and $\tau _{k}$
to $1$, it becomes clear that $S$ actually only depends on a single 
parameter ranging between $0$ and $1$. Explicitly, one has:

$$
S\left( \tau _{i},\tau _{j},\tau _{k}\right) =S\left( \frac{\tau _{ji}}{\tau
_{ki}}\right),
$$

\noindent
where, from now on, we use the notation 

$$
\tau _{ji}=\tau _{j}-\tau _{i}.
$$

\noindent
As we explicitly show in the appendix, the function $S\left( x\right) $ can
be computed exactly. It is a monotonically decreasing function defined on 
$\left[ 0,1\right] $ ranging from $1$ to $-1$. It 
satisfies $S\left( 1-x\right) =-S\left( x\right) $ and is explicitly 
given by

$$
S\left( x\right) =1-2L\left( x\right) .
$$

\noindent
The function $L\left( x\right) $ is the so called normalized Rogers
dilogarithm \cite{Rogers}, defined in terms of the usual dilogarithm 
$\mathrm{Li}_{2} \left( x\right) =\sum_{n=1}^{\infty }\frac{x^{n}}{n^{2}}$ 
as:

$$
L\left( x\right) =\frac{6}{\pi ^{2}}\left[ \mathrm{Li}_{2}\left( x\right) + 
\frac{1}{2}\ln \left( x\right) \ln \left( 1-x\right) \right] .
$$

\noindent
We then conclude that the contribution coming from the three--vertex 
$\mathcal{W}$ is given by 

$$
\sum_{i<j<k}S\left( \frac{\tau _{ji}}{\tau _{ki}}\right) W_{ijk}.
$$

\section{Computation of $n$--point Functions}

We now use the general results derived in the previous section in order 
to analyze the conformal properties of the $n$--point functions 
(\ref{pathInt}). Let us recall that we are still working in the simple 
case of constant symplectic structure $\omega _{ab}\left( x\right) 
=B_{ab}$, so that $\widetilde{\omega}_{ab} = B_{ab} + \frac{1}{3} 
H_{abc} x^{c}$. The generalization to arbitrary symplectic structure 
$\omega _{ab}\left( x\right) = B_{ab}+F_{ab}\left( x\right) $ is left to 
section 4.5.

In order to simplify the expressions in this section, we introduce the 
following short--hand notation:

\begin{eqnarray*}
K^{abc} &=&\theta ^{a\widetilde{a}}\theta ^{b\widetilde{b}} 
\theta^{c\widetilde{c}} H_{\widetilde{a}\widetilde{b}\widetilde{c}}\  \\
y_{a} &=&B_{ab}x^{b}.
\end{eqnarray*}

\subsection{2--point Function}

We shall first analyze the two--point function in some detail, since the 
manipulations for the higher point functions will be  similar. One 
considers two functions $f_{1}$ and $f_{2}$, placed at points $\tau _{1}$ and 
$\tau _{2}$ on the real line, with $\tau _{1}<\tau _{2}$. A simple computation, 
using the general results of the previous section, shows that the two--point 
function is explicitly given by:

\begin{equation}
\int f_{1}\star f_{2}+\int \left( -\frac{i}{6}K^{abc}{}\,y_{c}\star \partial
_{a}f_{1}\star \partial _{b}f_{2}+\mathcal{N}B_{bc}K^{abc}y_{a}\star
f_{1}\star f_{2}\right) .  \label{eq100}
\end{equation}

\noindent
The above expression does \textit{not} depend on the explicit values of 
$\tau_{1}$, $\tau_{2}$, but depends only on the order of the points 
$\tau _{i}$ on the real line. On the other hand, since two points on the 
boundary of a disk have no (conformal) moduli, the two--point function must 
be a \textit{symmetric} bilinear of $f_{1}$, $f_{2}$. The first term in 
(\ref{eq100}) is clearly symmetric. Let us then concentrate on the second 
term, by rewriting it with $f_{1}$ and $f_{2}$ interchanged. This gives, 
after a small rearranging,

\begin{equation}
\int \left( \frac{i}{6}K^{abc}{}\,\partial _{a}f_{1}\star y_{c}\star
\partial _{b}f_{2}+\mathcal{N}B_{bc}K^{abc}f_{1}\star y_{a}\star
f_{2}\right) .  \label{eq200}
\end{equation}

\noindent
Using that $K^{abc}{}\,\partial _{a}f_{1}\star y_{c}=K^{abc}{}\,y_{c}\star
\partial _{a}f_{1}$, and differentiating by parts, we see that the
difference between (\ref{eq200}) and the second term of (\ref{eq100}) 
reads 

$$
\int \left( \frac{i}{3}B_{bc}K^{abc}{}\,\partial _{a}f_{1}\star f_{2} 
- \mathcal{N}B_{bc}K^{abc}\left[ y_{a},f_{1}\right] \star f_{2}\right) .
$$

\noindent
The above is then vanishing if one has

$$
\mathcal{N}=\frac{1}{3}.
$$

\noindent
With this value the two--point function is conformally invariant and is a
simple symmetric bilinear of the functions $f_{1}$, $f_{2}$. Let us denote 
the $n$--point function by $P_{n}$. A little computation shows, using the
identity $\int f\star g=\int fg$, that the two--point function (\ref{eq100})
is given by the explicitly symmetric expression 

\begin{equation}
P_{2}\left( f_{1},f_{2}\right) =\int f_{1}f_{2}\left( 1+\frac{1}{3} 
H_{abc}x^{a}\theta ^{bc}\right) .  \label{eq400}
\end{equation}

\subsection{3--point Function}

Let $\tau_{1} < \tau_{2} < \tau_{3}$ be there \textit{ordered} points on 
the real line, and let us consider the three--point function of three 
functions $f_{1}$, $f_{2}$, $f_{3}$. One now has a contribution from the 
three--vertex $\mathcal{W}$, but it vanishes since

$$
K^{abc}\int \partial _{a}f_{1}\star \partial _{b}f_{2}\star \partial
_{c}f_{3}=0.
$$

\noindent
The only contribution then comes from the two--vertex $\mathcal{V}$, and 
it is given explicitly by (using the value $1/3$ for $\mathcal{N}$) 

\begin{eqnarray}
&&P_{3}\left( f_{1},f_{2},f_{3}\right) =\int f_{1}\star f_{2}\star f_{3} 
+ \frac{1}{3}B_{bc}K^{abc}\int y_{a}\star f_{1}\star f_{2}\star f_{3}
\nonumber \\
&&-\frac{i}{6}K^{abc}\int y_{c}\star \left( \partial _{a}f_{1}\star 
\partial_{b}f_{2}\star f_{3}+\partial _{a}f_{1}\star f_{2}\star 
\partial_{b}f_{3}+f_{1}\star \partial _{a}f_{2}\star \partial_{b} f_{3} 
\right). \label{eq300}
\end{eqnarray}

\noindent
As for the two--point function, the above expression does not depend on the
explicit values of the $\tau_{i}$'s, but only on their order on the real
line. On the other hand, since three points on the disk have no moduli (as
in the two--point function case), the above expression should actually be
invariant under cyclic permutations of the three functions, and in
particular under the replacement $f_{1}$, $f_{2}$, $f_{3}$ 
$\rightarrow$ $f_{2}$, $f_{3}$, $f_{1}$. One must then show that (\ref{eq300}) 
is equal to:

\begin{eqnarray*}
&&\int f_{2}\star f_{3}\star f_{1}+\frac{1}{3}B_{bc}K^{abc}\int f_{1}\star
y_{a}\star f_{2}\star f_{3} \\
&&-\frac{i}{6}K^{abc}\int \left( -\partial _{a}f_{1}\star y_{c}\star
\partial _{b}f_{2}\star f_{3}-\partial _{a}f_{1}\star y_{c}\star f_{2}\star
\partial _{b}f_{3}+f_{1}\star y_{c}\star \partial _{a}f_{2}\star \partial
_{b}f_{3}\right) .
\end{eqnarray*}

\noindent
In the second line of the above expression, we are free to move the
function $y_{c}$ all the way to the left, since we are contracting with the
totally antisymmetric object $K^{abc}$. Given this fact, it is simple to
show that the above expression is identical to (\ref{eq300}), thus proving
that also the three--point function is invariant under conformal
transformations, and is therefore a cyclic trilinear of its inputs. Note
that the \textit{same} value for $\mathcal{N}$ makes both $P_{2}$ and
$P_{3}$ invariant.

\subsection{4--point Function}

Let us now consider the four--point function. As usual we choose four
ordered points $\tau_{1} < \cdots < \tau_{4}$ on the real line and four
functions $f_{1}$, \dots, $f_{4}$. Following the general results in the
previous sections, the result of the path integral (\ref{pathInt}) breaks
into three parts. First we have the unperturbed result, given by

$$
\int f_{1} \star \cdots \star f_{4}.
$$

\noindent
The above is independent of the positions of the $\tau $'s, and is
conformally (actually topologically) invariant by itself, since it is a
cyclic multilinear function of the $f$'s. Second, we have the term coming
from the two--vertex $\mathcal{V}$, given by (in the notation of section 
3.2) 

\begin{equation}
V\left( f_{1}, \dots, f_{4}\right) =V+\sum_{i<j}V_{ij} . \label{eq800}
\end{equation}

\noindent
Finally we have, for the first time, a non--vanishing contributions to the
path integral coming from the three--vertex $\mathcal{W}$, which is given
explicitly by

\begin{equation}
\frac{1}{12}K^{abc}\int \left[ S\left( \tau _{1},\tau _{2},\tau _{3}\right)
\partial _{a}f_{1}\star \partial _{b}f_{2}\star \partial _{c}f_{3}\star
f_{4}+\cdots \right] ,  \label{eq700}
\end{equation}

\noindent
where $\cdots$ stands for three more terms which are weighted with the
corresponding factor $S\left( \tau _{i},\tau _{j},\tau _{k}\right)$, and
with the derivatives $\partial_{a}$, $\partial_{b}$ and $\partial _{c}$
acting on all possible groups of three functions ---as explained in 
section 3.2. Note that all the terms in (\ref{eq700}) are actually the same 
after integration by parts (for example $K^{abc}\int \partial _{a}f_{1}\star 
\partial _{b}f_{2}\star f_{3}\star \partial _{c}f_{4}=-K^{abc}\int 
\partial_{a}f_{1}\star \partial_{b}f_{2}\star \partial _{c}f_{3}\star 
f_{4}$, and so on), so that the above equation can be rewritten as 

\begin{equation}
\kappa \left( \tau _{i}\right) \frac{1}{12}K^{abc}\int f_{1}\star \partial
_{a}f_{2}\star \partial _{b}f_{3}\star \partial _{c}f_{4},  \label{eq900}
\end{equation}

\noindent
where the coefficient $\kappa$ is given by 

$$
\kappa \left( \tau _{i}\right) =-S\left( \tau _{1},\tau _{2},\tau
_{3}\right) +S\left( \tau _{1},\tau _{2},\tau _{4}\right) -S\left( \tau
_{1},\tau _{3},\tau _{4}\right) +S\left( \tau _{2},\tau _{3},\tau
_{4}\right) .
$$

Let us now discuss the conformal invariance of the above four--point
function. Start by considering a general $SL\left( 2,\mathbb{R}\right)$
transformation which \textit{preserves} the order of the points 
$\tau_{1}$, \dots, $\tau _{4}$, on the real line. In this case the term 
(\ref{eq800}) is invariant by itself, since it depends only on the order of 
the insertion points and not their specific positions. It must then be true 
that (\ref{eq900}) is also invariant, and this will be the case if 
the coefficient $\kappa\left( \tau _{i}\right) $ itself is unchanged under 
the $SL\left( 2,\mathbb{R}\right) $ transformation. We first recall that 
four points on the real line have a unique invariant module $m$, with 
$0<m<1$, which can be taken to be the position of point $2$ once one maps 
$\tau _{1}$, $\tau _{3}$, $\tau _{4}$ to $0$, $1$, $+\infty$. Using the 
standard notation $\tau _{ij}=\tau _{i}-\tau _{j}$, the module $m$ can also 
be invariantly described by the cross--ratio 

$$
m=\frac{\tau _{43}\tau _{21}}{\tau _{42}\tau _{31}}.
$$

\noindent
Let us now rewrite $\kappa $ in terms of Rogers dilogarithms (see the 
appendix)

$$
\frac{1}{2}\kappa =L\left( \frac{\tau _{21}}{\tau _{31}}\right) -L\left( 
\frac{\tau _{21}}{\tau _{41}}\right) +L\left( \frac{\tau _{31}}{\tau _{41}} 
\right) -L\left( \frac{\tau _{32}}{\tau _{42}}\right) .
$$

\noindent
If we use the general identity (\ref{app1}), from the appendix, with 
$x=\frac{\tau _{21}}{\tau_{31}}$ and $y=\frac{\tau _{31}}{\tau _{41}}$, 
one quickly discovers that 

$$
\kappa \left( \tau _{i}\right) =2L\left( m\right) ,
$$

\noindent
thus showing that the expression (\ref{eq700}) is conformally invariant, for 
an \textit{order--preserving} $SL\left( 2,\mathbb{R}\right)$ transformation.

One now needs to show that the full four--point function is invariant under
order--changing conformal transformations. We will actually be done once we
have considered the following special case. Start with the following
configuration of points $\tau_{1}=0$, $\tau_{2}=m$, $\tau_{3}=1$ and
$\tau_{4} = +\infty $. The $K$--dependent part of the four--point function
is given by

$$
V\left( f_{1},f_{2},f_{3},f_{4}\right) +\frac{1}{6}L\left( m\right)
K^{abc}\int f_{1}\star \partial _{a}f_{2}\star \partial _{b}f_{3}\star
\partial _{c}f_{4}.
$$

\noindent
Let us now move the point $\tau _{4}$ from $+\infty $ to $-\infty $. In this
case, the path integral gives 

\begin{eqnarray*}
&&V\left( f_{4},f_{1},f_{2},f_{4}\right) +\frac{1}{6}L\left( 1-m\right)
K^{abc}\int f_{4}\star \partial _{a}f_{1}\star \partial _{b}f_{2}\star
\partial _{c}f_{3} \\
&=&V\left( f_{4},f_{1},f_{2},f_{4}\right) +\frac{1}{6}\left( L\left(
m\right) -1\right) K^{abc}\int f_{1}\star \partial _{a}f_{2}\star \partial
_{b}f_{3}\star \partial _{c}f_{4}.
\end{eqnarray*}

\noindent
One then needs to prove that 

$$
V\left( f_{4},f_{1},f_{2},f_{4}\right) -V\left(
f_{1},f_{2},f_{3},f_{4}\right) =\frac{1}{6}K^{abc}\int f_{1}\star \partial
_{a}f_{2}\star \partial _{b}f_{3}\star \partial _{c}f_{4}.
$$

\noindent
We see that the dependence on the modulus $m$ has dropped out and this must
be the case since the LHS of the above equation depends only on the order of
the points, not on their positions. The above equation is a special case of
a more general formula which we shall prove in the next section, where we
consider the conformal invariance of $n$--point functions.

\subsection{General $n$--point Functions}

We finally turn our analysis to the $n$--point functions by considering the 
path integral with $n$ functions $f_{1}$, \dots, $f_{n}$ inserted on the real 
line in points $\tau _{1}<\cdots <\tau _{n}$ which are ordered from 
the left to the right. The unperturbed result is just,

$$
\int f_{1}\star \cdots \star f_{n},
$$

\noindent
which is invariant under all diffeomorphisms of the disk. The $H_{abc}$ 
dependent terms in the path integral divide as always in an expression 
coming from two--vertex,

\begin{equation}
V\left( f_{1}, \dots, f_{n}\right) =V+\sum_{i<j}V_{ij}\ , \label{eq2000}
\end{equation}

\noindent
and a part coming from the three--vertex $\sum_{i<j<k} S_{ijk} W_{ijk}$, 
where we compactly write $S_{ijk}=S\left(\tau_{i},\tau_{j},\tau_{k}\right)$.
We have defined the symbols $W_{ijk}$ and $S_{ijk}$ for $1\leq i<j<k \leq n$,
but one can extend the definition to all indices $i$, $j$, $k$ by demanding 
that both $W_{ijk}$ and $S_{ijk}$ be totally antisymmetric tensors.
Then the last contribution to the path integral is just

\begin{equation}
\frac{1}{6}\sum_{i,j,k}S_{ijk}W_{ijk}.  \label{eq1000}
\end{equation}

\noindent
The terms $W_{ijk}$ are not linearly independent since one can show,
differentiating by parts, that 

$$
\sum_{k}W_{ijk}=0.
$$

\noindent
This implies that\footnote{
These facts follow from the following (trivial) cohomology computation. 
Let $C^{k}$ be the space of totally antisymmetric tensors with $k$ indices, 
and let $\delta _{k+1}:C^{k+1}\rightarrow C^{k}$ be defined by $\delta
T_{i_{1}\cdots i_{k}}=\sum_{j}T_{i_{1}\cdots i_{k}j}$. Then $\delta
_{k+1}\delta _{k}=0$. It is easy to show that the corresponding cohomology
is trivial (see Equation \ref{qwer}).
Therefore if $\delta _{3}B=0$, it must be that $B=\delta
_{4}\cdots $. Moreover, one has that $\dim \ker \delta _{k+1}=\dim
C^{k+1}-\dim \func{Im}\delta _{k}=\dim C^{k+1}-\dim \ker \delta _{k}$, so
that $\dim \ker \delta _{3}=\dim C^{3}-\dim C^{2}+\dim C^{1}-\dim C^{0}$.}
the number of independent coefficients $W_{ijk}$ is $\left( ^{n-1} 
_{\;\;\, 3} \right)$, and that there is a totally antisymmetric tensor, 
$W_{ijkl}$, such that $W_{ijk}=\sum_{l}W_{ijkl}$. Concretely, one can choose 

\begin{equation}
W_{ijkl}=\frac{1}{n}\left( W_{ijk}-W_{ijl}+W_{ikl}-W_{jkl}\right).    \label{qwer}
\end{equation}

\noindent
Therefore equation (\ref{eq1000}) can be written as 

$$
\frac{1}{6}\sum_{i,j,k,l}S_{ijk}W_{ijkl}=\sum_{i<j<k<l}W_{ijkl}\left(
S_{ijk}-S_{ijl}+S_{ikl}-S_{jkl}\right)
$$

\noindent
As we have already seen in section 4.3 (see also the appendix), the 
properties of the Rogers dilogarithmic function imply that, for $i<j<k<l$, 

$$
S_{ijk}-S_{ijl}+S_{ikl}-S_{jkl}=-2L\left( \frac{\tau _{lk}\tau _{ji}}{\tau
_{lj}\tau _{ki}}\right) .
$$

\noindent
Therefore the final result,

\begin{equation}
-2\sum_{i<j<k<l}W_{ijkl}L\left( \frac{\tau _{lk}\tau _{ji}}{\tau _{lj}\tau
_{ki}}\right) ,
\end{equation}

\noindent
is written as a function only of the cross--ratios and is therefore
conformally invariant.

As in the case of the four--point function, the above reasoning is valid as
long as the conformal transformation preserves the order of the points on
the real line. In order to complete the proof of conformal invariance 
one must also consider the behavior of the full path integral as we pass
one point from $+\infty$ to $-\infty$. Let us then consider the simple
setup with $\tau_{1}$, \dots, $\tau_{n-1}$ at fixed positions, and $\tau
_{n}\rightarrow +\infty $. Then the sum (\ref{eq1000}) breaks into two 
parts:

$$
\sum_{i<j<k<n}S_{ijk}W_{ijk}+\sum_{i<j<n}S_{ijn}W_{ijn}=
\sum_{i<j<k<n}S_{ijk}W_{ijk}+\sum_{i<j<n}W_{ijn},
$$

\noindent
where we have used the fact that $S_{ijn}=S\left( 0\right) =1$. Let us now
``move $\tau_{n}$ across infinity'', so that $\tau _{n}\rightarrow -\infty 
$. The first term in the above expression is invariant, since it does not
contain the point $n$, and the function $f_{n}$ in $W_{ijk}$ is not
differentiated. The only change is in the term with $W_{ijn}$. As we move
the point $\tau _{n}$ from $+\infty $ to $-\infty $, the coefficients 
$W_{ijn}$ are multiplied not with $S\left( 0\right) =1$ but with $S\left(
1\right) =-1$, so that the total expression changes by

$$
2\sum_{i<j<n}W_{ijn}.
$$

\noindent
The above term is purely topological, \textit{i.e.}, it does not depend 
on the explicit position of the points $\tau_{1}$, \dots, $\tau_{n-1}$, 
and it must be canceled by the variation of expression (\ref{eq2000}) as we 
change the ordering of the functions. More precisely, one must have that:

$$
V\left( f_{n}, f_{1}, \dots, f_{n-1}\right) -V\left( f_{1}, \dots, 
f_{n}\right) =2\sum_{i<j<n}W_{ijn}.
$$

\noindent
To prove the above statement let us denote with $\widetilde{V}$ and 
$\widetilde{V}_{ij}$ the quantities corresponding to $V$ and $V_{ij}$, with
the functions $f_{1}$, \dots, $f_{n}$ permuted to $f_{n}$, $f_{1}$, 
\dots, $f_{n-1}$, so that $V\left( f_{n}, f_{1}, \dots, f_{n-1}\right) = 
\widetilde{V} + \sum_{i<j}\widetilde{V}_{ij}$. It is easy to show that,

\begin{eqnarray*}
\widetilde{V}_{ij} 
&=&2W_{i-1,j-1,n}+V_{i-1,j-1}\ ,\ \ \ \ \ \ \ \ \ \ \ \ \ \ 
\left( 1<i<j\right) \\
\widetilde{V}_{1j} &=&-V_{j-1,n}\ .\ \ \ \ \ \ \ \ \ \ \ \ \ \ \ 
\ \ \ \ \ \ \ \ \ \ \ \ \ \ \ \ \ \ \left( 1<j\right)
\end{eqnarray*}

\noindent
Also, since$\;\widetilde{V}-V=\frac{i}{3}B_{bc}K^{abc}\int f_{1}\star \cdots
\star f_{n-1}\star \partial _{a}f_{n}$ one can show that 

$$
\widetilde{V}-V=2\sum_{j<n}V_{jn}.
$$

\noindent
Putting everything together, one finally obtains 

\begin{eqnarray*}
&&V\left( f_{n}, f_{1}, \dots, f_{n-1}\right) -V\left( f_{1}, \dots, 
f_{n}\right) = \\
&=&\widetilde{V}-V+\sum_{1<i<j}\widetilde{V}_{ij}+\sum_{1<j}\widetilde{V}_{1j}
-\sum_{i<j<n}V_{ij}-\sum_{j<n}V_{jn} \\
&=&2\sum_{i<j<n}W_{ijn},
\end{eqnarray*}

\noindent
as was to be shown.

\subsection{Including the Boundary Interaction $S_{B}$}

In this section we are going to extend the results of the previous section
by including the effects of the boundary interaction $S_{B}$ in the
computation of the $n$--point functions (\ref{pathInt}). We have not checked
with path integral computations all the details of what follows, but the
extension is quite natural. We will leave for future work a detailed
path integral analysis of the results of this section.

It is natural in this context to change notation and to represent, as usual
(see, \textit{e.g.}, \cite{Cornalba-3}), functions as operators and $\star$
products with operator multiplication. Finally, integrals $\int$ will be
denoted by traces $\mathrm{Tr}$. Therefore, we shall shift notation for
functions as follows

$$
x^{a}\rightarrow X^{a}\ ,\ \ \ \ \ \ \ \ \ \ \ \ \ \ \ \ \ \ \ \
f_{i}\rightarrow F_{i}\ ,
$$

\noindent
and for traces as

$$
\int V\left( \omega \right) \,dx\rightarrow \mathrm{Tr}\ .
$$

\noindent
One then has the simple correspondences:

\begin{eqnarray*}
\theta ^{a\widetilde{a}}\partial _{\widetilde{a}}f &\rightarrow &-i [
X^{a},F ] , \\
\theta ^{ab} &\rightarrow &-i [ X^{a},X^{b} ] .
\end{eqnarray*}

\noindent
This allows us to rewrite the expressions for $V$, $V_{ij}$ and $W_{ijk}$ in
operator notation as

\begin{eqnarray*}
V &=&-\frac{2i}{3}H_{abc}\mathrm{Tr}\left( X^{a}X^{b}X^{c}F_{1}\cdots
F_{n}\right) , \\
V_{ij} &=&-\frac{i}{6}H_{abc}\mathrm{Tr}\left( X^{c}F_{1}\cdots [
X^{a},F_{i} ] \cdots [ X^{b},F_{j} ] \cdots F_{n}\right) ,
\end{eqnarray*}

\noindent
and,

$$
W_{ijk}=-\frac{i}{12}H_{abc}\mathrm{Tr}\left( F_{1}\cdots [ X^{a}, F_{i} ] 
\cdots [ X^{b},F_{j} ] \cdots [ X^{c}, F_{k} ] \cdots F_{n}\right) .
$$

We now consider the general case of $\omega_{ab}\left( x\right)
=B_{ab}+F_{ab}\left( x\right) $. The expressions above are still
well--defined and are the natural generalizations of the $\omega_{ab}
\left( x\right) =B_{ab}$ expressions previously derived. On the
other hand, for general $\omega $, we have that:

$$
\sum_{k}W_{ijk}=W_{ij},
$$

\noindent
where, for $i<j$, 

\begin{eqnarray*}
W_{ij} &=&\frac{i}{24}H_{abc}\mathrm{Tr}\left( F_{1}\cdots \left[ \left[
X^{a},X^{b}\right] ,F_{i}\right] \cdots \left[ X^{c},F_{j}\right] \cdots
F_{n}\right)  \\
&&-\frac{i}{24}H_{abc}\mathrm{Tr}\left( F_{1}\cdots \left[ X^{a},F_{i}\right]
\cdots \left[ \left[ X^{b},X^{c}\right] ,F_{j}\right] \cdots F_{n}\right) .
\end{eqnarray*}

\noindent
Note that, when $[ X^{a},X^{b} ] =i\theta ^{ab}$ is constant, $W_{ij}$ 
vanishes. In order to get a conformally invariant expression, one
is the forced to replace 

$$
W_{ijk}\rightarrow \frak{W}_{ijk}=W_{ijk}-\frac{1}{n}\left(
W_{ij}-W_{ik}+W_{jk}\right) .
$$

\noindent
It is then clear that $\sum_{k}\frak{W}_{ijk}=0$, so that the expression 
(using the notation of the previous sections) 

$$
\frak{W}=\sum_{i<j<k}S_{ijk}\frak{W}_{ijk}
$$

\noindent
is invariant under conformal transformations which do not change the order
of the insertion points on the real line. In the case analyzed in the 
previous section, the term above (coming from the three--vertex) was 
supplemented with the term coming from the two--vertex,

$$
V\left( F_{1}, \dots, F_{n}\right) =V+\sum_{i<j}V_{ij}.
$$

\noindent
We recall that the above expression is important in the case when $\tau_{n}$
``goes around $\infty$''. In particular, when $[ X^{a},X^{b} ]$
is constant, we have that $V\left( F_{n},F_{1},\dots,F_{n-1} \right)
-V\left( F_{1},\dots,F_{n} \right) =2\sum_{i<j<n}\frak{W}_{ijn}$, so that
the full $n$--point function is conformally invariant. Again, for general 
$[ X^{a},X^{b} ]$, we must add to $V\left( F_{1},\dots,F_{n} \right) $ 
terms which vanish for constant $[ X^{a},X^{b} ]$. The simplest way to find 
the correct result is the following. First let us introduce a bit of notation. 
As in section 4.4, given any expression $\cdots$, we will denote with 
$\widetilde{\cdots}$ the same expression, with the functions $F_{1}$, \dots, 
$F_{n}$ cyclically permuted to $F_{n}$, $F_{1}$, \dots, $F_{n-1}$. In 
particular, for $1<i<j<k$, one has that 
$\widetilde{\frak{W}}_{ijk}=\frak{W}_{i-1,j-1,k-1}$ and that 
$\widetilde{\frak{W}}_{1ij}=\frak{W}_{i-1,j-1,n}$. Let us then consider the 
following expression:

$$
v\left( F_{1}, \dots, F_{n}\right) =\frac{2}{n}\sum_{i<j<k}\left(
i+j+k\right) \frak{W}_{ijk} .
$$

\noindent
A small computation shows that 

\begin{eqnarray*}
\widetilde{v} &=&\frac{2}{n}\sum_{i<j<k<n}\left( i+j+k+3\right) 
\frak{W}_{ijk}+\frac{2}{n}\sum_{i<j<n}\left( i+j+3\right) \frak{W}_{ijn} \\
&=&v-2\sum_{i<j<n}\frak{W}_{ijn},
\end{eqnarray*}

\noindent
where we have used the fact that $\sum_{i<j<k}\frak{W}_{ijk}=0$, which
follows simply from $\sum_{k}\frak{W}_{ijk}=0$. One can then consider the
combination:

\begin{equation}
V\left( F_{1}, \dots, F_{n}\right) +v\left( F_{1}, \dots, F_{n}\right) .
\label{V-corr}
\end{equation}

\noindent
The previous discussion implies that expression (\ref{V-corr}), in the case
of constant $[ X^{a},X^{b} ] =i\theta^{ab}$, is a \textit{cyclic} 
function in the arguments $F_{1}$, \dots, $F_{n}$. In general, though, the
above need not be cyclic. We can nonetheless construct the correct
generalization, $\frak{V}$ of $V\left( F_{1},\dots,F_{n} \right) $, by
cyclically symmetrizing. In particular, if we define

$$
\frak{V}=\frac{1}{n}\left[ V\left( F_{1},\dots,F_{n} \right) +v\left(
F_{1},\dots,F_{n} \right) +\mathrm{cyclic}_{1\cdots n}\right] -v\left(
F_{1},\dots,F_{n} \right) ,
$$

\noindent
then this satisfies

$$
\widetilde{\frak{V}}\frak{-V}=2\sum_{i<j<n}\frak{W}_{ijn}
$$

\noindent
and, following the same arguments as in section 4.4, we have restored
conformal invariance. Therefore the final result for the $n$--point function
is given by:

$$
\frak{V}+\sum_{i<j<k}S_{ijk}\frak{W}_{ijk}.
$$

\section{Nonassociative Deformations of Worldvolumes}

We are now in a position to show the importance of the Kontsevich product
$\bullet$ in the above construction. In this section we shall first discuss
in some detail Kontsevich products defined starting from various different
bi--vector fields (in section 5.1), and then see how one can reinterpret,
in this framework, the results of the last section (in 5.2 and 5.3).

Let us start the discussion by considering the simplest case when $\omega
=B+F$ is constant. We are then considering the standard Moyal product
deformation of the brane world--volume, which is described in \cite
{Schomerus, Seiberg-Witten}. Physically, it corresponds to the embedding of
a flat brane in a flat background space. The relevant product is the Moyal
star product, given by the formula

\begin{equation}
\left( f\star g\right) \,\left( x\right) =e^{{\frac{i}{2}}\theta
^{ij}\partial _{i}^{x}\partial _{j}^{y}}\ f(x)g(y)|_{x=y}.
\end{equation}

\noindent 
The open string parameters can be written in terms of the closed string
parameters with the formulas (\ref{open-parameters}), where $\theta^{ab} =
-i[x^{a},x^{b}]_{\star}$. In the zero slope limit \cite {Seiberg-Witten},
correlators are computed according to:

$$
\left\langle \prod_{i=1}^{n} f_{i} \left( X(\tau_{i}) \right) \right\rangle
= \int \sqrt{\det \omega}\ d^{p+1}x\ f_{1} \star \cdots \star f_{n}.
$$

Now let us consider the case when $\omega \left( x\right)$ is no longer
constant, but $d\omega =0$. Then, $\omega $ still defines a symplectic
structure on the brane world--volume. Physically, this corresponds to
embeddings of a curved brane in a flat background space, as can be most
easily seen from the Matrix theory point of view (for example this is
described, in the context of holomorphic curves in flat space, in
\cite{Cornalba-Taylor, Cornalba-1}). Recall, in fact, that the $F$ field
represents the zero--brane density on a two--brane, such that

$$
N = {\frac{1}{2\pi}} \int_{S} F
$$

\noindent 
is the total number of zero--branes. For static solutions $F$ is
proportional to the area element, and is therefore no longer constant with
respect to the Euclidean coordinates of the flat background. The
zero--brane density varies along the two--brane, which in turn effectively
amounts to building a curved $M2$--brane in the flat space background.

From the $\sigma$--model point of view, the case of $d\omega =0$ is very
similar to the constant one (after all, all symplectic structures are
locally related by a coordinate change), but now the Moyal star product is
replaced by Kontsevich's formula \cite{Kontsevich}, as shown in detail in
\cite{Cattaneo-Felder}. Then the star product is (we denote the
noncommutative parameter by $\alpha^{ab} \left( x \right)$ in here),

\begin{eqnarray}
f\star g &=&fg+{\frac{i}{2}}\alpha ^{ab}\partial _{a}f\partial _{b}g-
{\frac{1}{8}}\alpha ^{ac}\alpha ^{bd}\partial _{a}\partial _{b}f
\partial_{c}\partial _{d}g  \notag \\
&&-{\frac{1}{12}}\alpha ^{ad}\partial _{d}\alpha ^{bc}\left( \partial
_{a}\partial _{b}f\partial _{c}g-\partial _{b}f\partial _{a}\partial
_{c}g\right) +\mathcal{O}(\alpha ^{3}),  \label{kontsevich}
\end{eqnarray}

\noindent 
while open string parameters are still given by the same formulas.
Finally, correlators are computed in the $\alpha^{\prime} \rightarrow 0$
limit as follows,

$$
\left\langle \prod_{i=1}^{n} f_{i} \left( X(\tau_{i}) \right) \right\rangle
= \int V(\omega) d^{p+1}x\ \left( f_{1} \star \cdots \star f_{n} \right).
$$

The situation we analyze in detail in this paper is when $H = d\omega \neq
0$.  The target is then no longer flat and one is thus embedding a curved
brane in a curved background. At first sight it seems that, since one no
longer has a symplectic manifold (and therefore a Poisson structure), one
can no longer identify the correct algebraic structure ---if any--- which
controls the deformation in this case. The result we have obtained is that
this is not the case. As we will explain at length in this section, the
Kontsevich formula is still relevant in the description of the
physics. Indeed, we find that the deformation is \textit{still} given by
the Kontsevich star product expansion, as written in coordinates (we shall
now denote the star product by $\bullet$ and the inverse two--form by
$\widetilde{\alpha}$ in order to distinguish the two cases),

\begin{eqnarray}
f\bullet g &=&fg+{\frac{i}{2}}\widetilde{\alpha }^{ab}\partial _{a}f\partial
_{b}g-{\frac{1}{8}}\widetilde{\alpha }^{ac}\widetilde{\alpha }^{bd}\partial
_{a}\partial _{b}f\partial _{c}\partial _{d}g  \notag
\label{kontsevich-nonassociative} \\
&&-{\frac{1}{12}}\widetilde{\alpha }^{ad}\partial _{d}\widetilde{\alpha}^{bc}
\left( \partial _{a}\partial _{b}f\partial _{c}g-\partial _{b}f\partial
_{a}\partial _{c}g\right) +\mathcal{O}(\widetilde{\alpha }^{3}).
\end{eqnarray}

\noindent 
The difference now is that the star product is no longer
associative. Therefore, when in curved backgrounds, the brane world--volume
is deformed not only through a noncommutative parameter
($\widetilde{\alpha} = \widetilde{\omega }^{-1}$), but also through a
nonassociative parameter which ---as we shall see--- is essentially
$H=d\widetilde{\omega }$.

Again, open string parameters are given by the same formulas as in the
Moyal case (as will be later shown in section 6). As we have seen in the
previous section, correlators in the topological limit $g_{ab} \rightarrow
0$ require a detailed analysis. The results can again be written, as we
shall show in this section, in terms of $\bullet$ and the general formula
will still be

$$
\left\langle \prod_{i=1}^{n}f_{i}\left( X(\tau _{i})\right) \right\rangle
\sim \int V(\widetilde{\omega })d^{p+1}x\ \left( f_{1}\bullet \cdots \bullet
f_{n}\right) .
$$

\noindent 
On the other hand, due to the non--associativity of $\bullet$, one has to
define precisely what one means by the RHS of the above equation, which now
depends explicitly on the moduli of the insertion points $\tau_{i}$.

\subsection{Nonassociative Star Products}

We shall now study the properties of the nonassociative Kontsevich star
product $\bullet$. In the last part of this section we will work in the
$g_{ab} \rightarrow 0$ limit, so that $\alpha = \omega^{-1}$,
$\widetilde{\alpha} = \widetilde{\omega}^{-1}$. Also, unless explicitly
needed, we shall drop the tildes.

Let us start by considering the associativity properties of the Kontsevich
expansion (\ref{kontsevich}) or (\ref{kontsevich-nonassociative}). To this
end one needs to compute, given three generic functions $f$, $g$, $h$, the
difference $(f\star g)\star h-f\star (g\star h)$. Using the expansions
(\ref{kontsevich}), (\ref{kontsevich-nonassociative}), it is not difficult
to show that:

\begin{equation}  \label{associative}
\left( f \star g \right) \star h - f \star \left( g \star h \right) = 
\frac{1}{6} \left( \alpha^{i\ell} \partial_{\ell} \alpha^{jk} + \alpha^{j\ell}
\partial_{\ell} \alpha^{ki} + \alpha^{k\ell} \partial_{\ell} \alpha^{ij}
\right) \left( \partial_{i} f \partial_{j} g \partial_{k} h \right) + 
\mathcal{O}(\alpha^{3}).
\end{equation}

\noindent 
If $\alpha $ is a Poisson structure, \textit{i.e.}, satisfies

$$
\alpha^{i\ell} \partial_{\ell} \alpha^{jk} + \alpha^{j\ell} \partial_{\ell}
\alpha^{ki} + \alpha^{k\ell} \partial_{\ell} \alpha^{ij} = 0,
$$

\noindent 
then the associated product is associative (in fact to all orders). Note
that, when $\alpha$ is invertible, the above equation is equivalent to
$d\left( \alpha^{-1} \right) = d\omega = 0$, so that the expansion
(\ref{kontsevich}) defines an associative product.

Now consider the product (\ref{kontsevich-nonassociative}). In this case
one has both a noncommutative deformation, with parameter
$\widetilde{\alpha}$, and a nonassociative deformation with parameter $H =
d\widetilde{\omega}$. To better understand this point let us re--write
expression (\ref{associative}) in terms of the 3--form field $H$. Indeed,
using that $\partial_{k} \widetilde{\alpha}^{ij} = \widetilde{\alpha}^{ia}
\widetilde{\alpha}^{jb} \partial_{k} \widetilde{\omega}_{ab}$, one can
rewrite (\ref{associative}) as

\begin{equation}  \label{nonassociative}
\left( f \bullet g \right) \bullet h - f \bullet \left( g \bullet h \right)
= \frac{1}{6}\, \widetilde{\alpha}^{ia} \widetilde{\alpha}^{jb}
\widetilde{\alpha}^{kc}\, H_{abc}\, \partial_{i}
f\, \partial_{j} g\, \partial_{k} h + \cdots .
\end{equation}

\noindent 
Precisely because we have $H\not=0$, the star product (\ref
{kontsevich-nonassociative}) is \textit{not} associative.

We then have two products, $\star$ and $\bullet$, given by the Kontsevich
expansion in terms of $\omega_{ab} \left( x\right)$ and
$\widetilde{\omega}_{ab} \left( x\right) = \omega_{ab} \left( x\right) +
\frac{1}{3}H_{abc}x^{c}$, respectively. We wish to explicitly relate the
product $\bullet$ to the associative product $\star$. First, it is clear
that:

$$
\widetilde{\alpha }^{ij}=\alpha ^{ij}+\frac{1}{3}\alpha ^{ai}\alpha
^{bj}H_{abc}x^{c} + \cdots .
$$

\noindent
Therefore one has $f \bullet g = f \star g + \frac{i}{6} \alpha^{ai}
\alpha^{bj} H_{abc} x^{c} \partial_{i} f \partial_{j} g +
\cdots$. Recalling that $\left[ x^{a},f \right] = i\alpha^{ai} \partial_{i}
f + \cdots$, it is not hard to show:

$$
f\bullet g = f\star g - \frac{i}{12} H_{abc} \left\{ x^{c}, [ x^{a},f
]_{\star} \star [ x^{b},g ]_{\star} \right\}_{\star}  .
$$

\noindent
One can check the correctness of the above formula by expanding equation
(\ref{kontsevich-nonassociative}) to order $\alpha^{2}$. It is moreover
convenient, as in section 4.5, to move to operator notation for the
associative product $\star$. Therefore the above expression for the product
$\bullet$ can be compactly written as

\begin{equation} \label{FbulletG}
F\bullet G = FG - \frac{i}{12} H_{abc} \left\{ X^{c}, [ X^{a},F ] [ X^{b},G
] \right\} .
\end{equation}

\noindent
It is then simple to show that:

$$
\left( F\bullet G\right) \bullet H - F\bullet \left( G\bullet H\right) = - 
\frac{i}{6}\ H_{abc}\ [ X^{a},F ] [ X^{b},G ] [ X^{c},H ] ,
$$

\noindent
which is the generalization of (\ref{nonassociative}) to all orders in
$\alpha$.

Let us now take the functions $f$, $g$ and $h$ to be the local coordinate
functions $x^{i}$ in $\mathbb{R}^{n}$. By direct use of the nonassociative
Kontsevich formula (\ref{kontsevich-nonassociative}) one obtains,

\begin{equation}  \label{product-coordinates}
x^{i} \bullet x^{j} = x^{i} x^{j} + {\frac{i}{2}} \widetilde{\alpha}^{ij}(x),
\end{equation}

\noindent 
and from (\ref{nonassociative}),

\begin{equation}  \label{nonassociative-coordinates}
\left( x^{i} \bullet x^{j} \right) \bullet x^{k} - x^{i} \bullet \left(
x^{j} \bullet x^{k} \right) = \frac{1}{6}\, \widetilde{\alpha}^{ia} 
\widetilde{\alpha}^{jb} \widetilde{\alpha}^{kc}\, H_{abc} .
\end{equation}

\noindent 
Calculating the star bracket commutator (making use of
(\ref{product-coordinates})) one obtains the noncommutative algebra,

$$
[x^{i}, x^{j}]_{\bullet} = i \widetilde{\alpha}^{ij}(x),
$$

\noindent 
which is a very similar result to the standard Kontsevich deformation. On
the other hand, in order to compute the Jacobi expression, one uses
(\ref{nonassociative-coordinates}) to obtain:

$$
[x^{i}, [x^{j}, x^{k}]_{\bullet}]_{\bullet} + [x^{j}, [x^{k},
x^{i}]_{\bullet}]_{\bullet} + [x^{k}, [x^{i}, x^{j}]_{\bullet}]_{\bullet} =
- \widetilde{\alpha}^{ia} \widetilde{\alpha}^{jb} \widetilde{\alpha}^{kc}\, H_{abc},
$$

\noindent 
which is a violation of the Jacobi identity.

\subsection{Operator Product Expansions and Factorization}

In the previous subsection we have reviewed the basic properties of the
nonassociative Kontsevich product $\bullet$. We may now use the general
results of section 4.5 in order to show the relevance of the algebraic
operation $\bullet$ in the computation of $n$--point functions. Let us then
first summarize the results of section 4. We have constructed $n$--point
functions $P_{n} \left[ F_{1}, \dots, F_{n} \right]$ which depend uniquely
on the $n-3$ conformal moduli of the insertion points $\tau_{i}$ of the
functions $F_{i}$. In particular, the one--point function $P_{1}$, which we
shall call $P$ in the sequel, is a generalization of the trace,
$\mathrm{Tr}$, and is given by:

$$
P_{1}\left[ F\right] =P\left[ F\right] =\mathrm{Tr}\left( F\right) - 
\frac{2i}{3}H_{abc}\mathrm{Tr}\left( X^{a}X^{b}X^{c}F\right) .
$$

\noindent
Now consider a general $n$--point function for functions $F_{i}$ at points
$\tau_{i}$. Let us scale the insertion points $\tau_{i} \rightarrow
\varepsilon \, \tau_{i}$ for $\varepsilon \rightarrow 0$. On one hand the
result of the $n$--point function does not change, since it is invariant
under $SL\left( 2,\mathbb{R}\right)$ transformations. On the other hand we
can use an OPE argument to conclude that there must exist a function,
$O_{n} \left[ F_{1}, \dots, F_{n} \right] \left( \tau_{i}\right)$, such
that

\begin{equation} \label{PfromO}
P_{n}\left[ F_{1}, \dots, F_{n}\right] \left( \tau _{i}\right) = P \left[
O_{n} \left[ F_{1}, \dots, F_{n}\right] \left( \tau _{i}\right) \right].
\end{equation}

\noindent
The operations $O_{n} \left[ F_{1}, \dots, F_{n} \right] \left( \tau_{i}
\right)$ are ---informally--- untraced versions of the $P_{n}$'s, and are
invariant under the subgroup of $SL\left( 2,\mathbb{R}\right)$ which leaves
the point at $\infty$ invariant, \textit{i.e.}, translations and
rescalings. They will depend on $n-2$ moduli. In particular one can now see
the relevance of the $\bullet$ product, which is nothing but the operation
$O_{2}$. More precisely, one can check that

\begin{eqnarray*}
O_{1}\left[ F\right] \left( \tau \right) &=&F , \\
O_{2}\left[ F,G\right] \left( \tau _{1},\tau _{2}\right) &=& F\bullet G ,
\end{eqnarray*}

\noindent
where, for the second expression, it is simple to use its explicit
expansion, (\ref{FbulletG}), and insert it in (\ref{PfromO}) in order to
check that one does get the right result, $P_{2} [F,G]$, as derived in
section 4.5. With a little more work, and using the results on the
$n$--point functions of section 4 and the facts on the $\bullet$ product
of the previous subsection, one can likewise obtain,

\begin{eqnarray*}
O_{3}\left[ F,G,H\right] \left( \tau _{i}\right) &=&L\left( 1-m\right)
\,\left( F\bullet G\right) \bullet H+L\left( m\right) \,F\bullet \left(
G\bullet H\right) , \\
m &=&\frac{\tau _{21}}{\tau _{31}}.
\end{eqnarray*}

\noindent
In particular $O_{3}$, which depends on a single modulus, is explicitly
written in terms of the product $\bullet$ and interpolates between the two
possible positionings of the parenthesis. More generally, the operations
$O_{n}$ will depend on $n-2$ moduli and will interpolate between the
various possible ways of taking products of $n$ functions with the
$\bullet$ product.

OPE arguments can be used, in the general case, to compute $n$--point
functions at the boundary of the moduli space of the insertion points
$\tau_{i}$.  Again consider an $n$--point function with functions $F_{i}$
at points $\tau_{i}$. Let a subset of the points ---say $\tau_{1}$,
$\dots$, $\tau_{m}$--- converge to zero via a common rescaling $\tau_{i}
\rightarrow \varepsilon \tau_{i}$, $i=1$, $\dots$, $m$. Then one can use an
OPE argument to show that (the indices $i$ and $j$ indicate the two sets
$1$, $\dots$, $m$, and $m+1$, $\dots$, $n$, respectively)

\begin{equation} \label{Plimit}
\lim_{\varepsilon \rightarrow 0}P_{n}\left[ F_{1}, \dots, F_{n}\right]
\left( \varepsilon \tau _{i},\tau _{j}\right) =P_{n-m+1}\left[ O_{m}\left[
F_{1}, \dots, F_{m}\right] \left( \tau _{i}\right), F_{m+1}, \dots, F_{n} 
\right] \left( 0,\tau _{j}\right) .
\end{equation}

\noindent
For example, one can show that:

$$
P\left[ \left( F\bullet G\right) \bullet H\right] =P\left[ F\bullet \left(
G\bullet H\right) \right] .
$$

\noindent
This follows from applying (\ref{Plimit}) and recalling the fact that the
three--point function $P_{3} \left[ F,G,H \right] \left( \tau_{1},
\tau_{2}, \tau_{3} \right)$ is independent of the moduli. If one considers
the two limits $\tau_{2} \rightarrow \tau_{1}$ and $\tau_{2} \rightarrow
\tau_{3}$, and uses factorization with $O_{2} \sim \bullet$, one quickly
arrives at the above result.

\subsection{The Homotopy Associative Algebraic Structure}

We have seen that the operations $O_n$ define, as a function of the modular
parameters, a structure which extends that of an associative algebra. In
fact, the failure of the product $\bullet \sim O_2$ to be associative is
measured by $O_3$, which now interpolates (thanks to the modular parameter
$0 < m < 1$) between the two possible ``placements'' of the parenthesis. In
a very crude sense, the nonassociativity at each order is controlled by
higher order terms. These type of structures have appeared in the
literature on string theory, starting from the use in string field theory
of the Batalin--Vilkovisky formalism to quantize gauge theories which do
not close off-shell \cite{Lada-Stasheff, Lada-Markl, Zwiebach, GAB}. These 
are the $A_{\infty}$ homotopy associative algebras, where the failure of 
the associativity property is controlled by a third order term, and similarly  
at higher orders \cite{Stasheff}.

Let us formalize these concepts a bit further, and show that the 
structure $C^{\infty}(M)$ and $O_{n}[F_{1}, \ldots, F_{n}](\tau)$ is 
actually that of an $A_{\infty}$ space\footnote{In here we take $M$ to 
be the brane world--volume.}. The idea of an $A_{\infty}$ 
space is the same as that of an $A_{\infty}$ algebra, only the definition 
of homotopy is changed (one uses a map instead of a differential). So 
we first follow the original work of Stasheff \cite{Stasheff} and 
recall the definition of homotopy associativity. The intuitive notion 
is the following. A space $X$ and a multiplication $m : X \times X \to 
X$ is a homotopy associative space if the maps $m(\mathbf{1} \otimes 
m)$ and $m(m \otimes \mathbf{1})$ are homotopic as maps $X \times X 
\times X \to X$. If we are given three functions, $F_{1}$, $F_{2}$ 
and $F_{3}$, with the nonassociative product $\bullet$ there are two 
distinct ways to insert parenthesis in the natural application 
$C^{\infty}(M) \times C^{\infty}(M) \times C^{\infty}(M) \to 
C^{\infty}(M)$, \textit{i.e.}, the standard options $\left( F_{1} 
\bullet F_{2} \right) \bullet F_{3}$ and $F_{1} \bullet \left( F_{2} \bullet 
F_{3} \right)$. But a quick reminder of the previous section also tells us 
that there is a homotopy, $O_{3} [F_{1}, F_{2}, F_{3}] (m) : [0,1] \times 
C^{\infty}(M)^{\times 3} \to C^{\infty}(M)$, between these two seemingly 
distinct ways to associate brackets under the $\bullet$ product.

In order to realize that there are stronger conditions of 
``associativity modulo homotopy'' than the previous one, let us 
proceed by analyzing the situation with four functions, $F_{1}, 
\ldots, F_{4}$. There are now five distinct ways to insert 
parenthesis for the nonassociative product,
\begin{eqnarray*}
&( F_{1} \bullet F_{2} ) \bullet ( F_{3} \bullet F_{4} ),& \\
&( ( F_{1} \bullet F_{2} ) \bullet F_{3} ) \bullet F_{4}, \;\;\;\;\;\; 
\;\;\;\;\;\; \;\;\;\;\;\; \;\;\;\;\;\; \;\;\;\;\;\;
F_{1} \bullet ( F_{2} \bullet ( F_{3} \bullet F_{4} ) ),& \\
&( F_{1} \bullet ( F_{2} \bullet F_{3} ) ) \bullet F_{4}, \;\;\; 
\;\;\; \;\;\;
F_{1} \bullet ( ( F_{2} \bullet F_{3} ) \bullet F_{4} ),&
\end{eqnarray*}
\noindent
which can actually be pictorially written at the vertices of a pentagon. 
The point is now that while the $O_{3}$ homotopy naturally yields 
homotopies that run between the vertices, it is not necessarily true 
that one can extend the homotopy to the interior of the pentagon. If 
one can \textit{not} extend the homotopy to this situation, the algebraic 
structure of the product is denoted $A_{3}$. If, on the other hand, 
one \textit{can} extend the homotopy to the interior of the whole pentagon, 
the algebraic structure is denoted $A_{4}$. As we go further along this way 
one is led to consider higher poliedra, and if one can always extend 
homotopies to the interior of these poliedra, then the algebraic 
structure is $A_{\infty}$ homotopy associative \cite{Stasheff}.

Let us illustrate these concepts by explictly writing down the 
$O_{4}[F_{1}, \ldots, F_{4}](x,y)$ operation. One can compute it to be

\begin{eqnarray*}
O_{4}[F_{1}, \ldots, F_{4}](x,y) &=& L \left[ \left( 1 - \frac{x}{y} 
\right) \left( 1 - \frac{1-y}{1-x} \right) \right] ( F_{1} \bullet F_{2} ) 
\bullet ( F_{3} \bullet F_{4} )\\
&&+ L \left[ \left( 1 - \frac{x}{y} \right) \left( \frac{1-y}{1-x} \right) 
\right] ( ( F_{1} \bullet F_{2} ) \bullet F_{3} ) \bullet F_{4}\\
&&+ L \left[ \frac{x}{y} \left( 1 - \frac{1-y}{1-x} \right) \right] F_{1} 
\bullet ( F_{2} \bullet ( F_{3} \bullet F_{4} ) )\\
&&+ L \left[ \frac{x}{y} \left( 1 - y \right) \right] ( F_{1} \bullet ( 
F_{2} \bullet F_{3} ) ) \bullet F_{4}\\
&&+ L \left[ x \left( \frac{1-y}{1-x} \right) \right] F_{1} \bullet ( ( 
F_{2} \bullet F_{3} ) \bullet F_{4} ), \\
x &=& \frac{\tau_{21}}{\tau_{41}}\ , \;\;\;\;\;\; y = 
\frac{\tau_{31}}{\tau_{41}},
\end{eqnarray*}

\noindent
where $0 < x < y <1$. At first sight one would say that $\{x,y\}$ take 
values in a triangle. However, a glance at the expression above also 
tells us that while one of the vertices of this triangle, 
$\{x,y\}=\{0,1\}$, is perfectly regular, the other two, 
$\{x,y\}=\{0,0\}$ and $\{x,y\}=\{1,1\}$, are actually singular. Each 
of these singular points can actually be resolved into two distinct 
limits, once we scale $x$ and $y$ in the two possible different ways. 
For instance, the limit where $\{x,y\} \to \{0,0\}$ can be approached 
with both $x$ and $y$ scaling as $\epsilon \to 0$, with $\frac{x}{y} 
\to 1$ or with $x$ scaling as $\epsilon^{2}$ and $y$ as $\epsilon$, 
and $\frac{x}{y} \to 0$. A similar situation occurs for the limit 
where $\{x,y\} \to \{1,1\}$. So, the resolution of the singular vertices 
of the triangle actually produces the expected pentagon. Once this is 
realized, it is simple to see that $O_{4}$ plays the role of the $A_{4}$ 
homotopy,

$$
O_{4}[F_{1}, \ldots, F_{4}](x,y) : {\cal P} \times 
C^{\infty}(M)^{\times 4} \to C^{\infty}(M),
$$

\noindent
where ${\cal P}$ is the pentagon spanned by $x$ and $y$. Observe that, 
as explained in the previous section, $P[O_{4}]=O_{4}$. It is then very 
natural to conjecture that general $n$--point functions will thus 
produce the necessary homotopies in order that $\left( 
C^{\infty}(M), \bullet, \{ O_{n} \}_{n=2}^{\infty} \right)$ is an 
$A_{\infty}$ homotopy associative algebra (and where $P_{n} = 
P[O_{n}]$).

The homotopies $O_{n}[F_{1}, \ldots, F_{n}](\tau)$ also induce
the necessary homotopies to create an $L_{\infty}$ commutator homotopy Lie
algebra, and to create homotopy differential operators (which may contain 
non--trivial topological information for the tensor bundles of $M$). 
Indeed, the commutator algebra, $[x^{i}, x^{j}]_{\bullet}$, 
is an $L_{\infty}$ homotopy Lie algebra: using the basic homotopy, $O_{3} 
[F,G,H](m)$, one can define a ``composite'' homotopy between zero and  
$-\frac{i}{6} H_{abc} [X^{a},F] [X^{b},G] [X^{c},H]$, the term that violates 
Jacobi's identity in the $\bullet$ commutator algebra. With this 
homotopy, Jacobi's identity will be satisfied up to homotopy, and one thus 
obtains an $L_{\infty}$ homotopy Lie algebra. In order to build a differential 
structure (and thus, gauge theory) one still needs a covariant derivative in 
the sense that $\nabla \left( F \bullet G \right) = \nabla F \bullet G + F 
\bullet \nabla G$. While this may not seem as a viable course of action, one 
can use \textit{homotopy} to impose the Leibnitz rule: the derivative 
operation $\nabla_{X} F \sim [ X, F ]$ will satisfy the Leibnitz rule 
\textit{up to} homotopy. Again using the basic homotopy one can define a 
composite homotopy between $[ X, F \bullet G ]$ and $[X,F] \bullet G + F 
\bullet [X,G]$, so that the commutator $[X,F]$ becomes a homotopy derivative 
for the $\bullet$ product.

\section{Corrections Involving the Metric Tensor}

Up to now we have studied the topological limit $g_{ab}\rightarrow 0$ in
great detail. In these last sections we shall discuss corrections to the
above results, when one includes a non--vanishing closed string metric
$g_{ab}$ in the calculations. This will allow us to identify the open
string effective parameters ---metric $G^{ab}$ and noncommutative parameter
$\theta^{ab}$--- in terms of the closed string parameters $g_{ab}$ and
$B_{ab}$.

Recall that the two--point function of the fluctuating field $\zeta$ is

$$
\langle \zeta^{a} \left( z \right) \zeta^{b} \left( \tau \right) \rangle = 
\frac{i}{\pi} \theta^{ab} \mathcal{A} \left( z,\tau \right) - \frac{1}{\pi}
G^{ab} \mathcal{B} \left( z,\tau \right),
$$

\noindent 
where

\begin{eqnarray*}
\mathcal{A} \left( z,\tau \right) &=& \frac{1}{2i} \ln \left( \frac{\tau - 
z}{\overline{z} - \tau} \right), \\
\mathcal{B} \left( z,\tau \right) &=& \ln \left|z - \tau \right|,
\end{eqnarray*}

\noindent 
and where we have placed the point $\tau$ at the world--sheet boundary
$\partial\Sigma$. In the previous sections we have worked only with the
term in $\mathcal{A}(z,\tau)$. Here, we will evaluate the two--point
function $P_{2} (x^{a}, x^{b})$ including the contribution arising from the
term in $\mathcal{B}(z,\tau)$. The only diagrammatic contribution still
comes from the two--graph $\mathcal{V}$, (\ref{V2}), so that one can easily
compute the relevant Feynman diagrams. First, observe that the contribution
of the $\mathcal{B}(z,\tau)$ term to the volume form $V(\omega)$ is
proportional to $\propto H_{abc} G^{ab}$ and therefore vanishes due to the
antisymmetry of $H_{abc}$. Thus, the volume form is unchanged.

Schematically, the propagator looks like $\mathcal{A}-\mathcal{B}$. In the
previous sections we computed the correction to the two--point function
going like $\mathcal{A}^{2}$, with the result:

$$
{\frac{i }{3\pi}} \theta^{ia} \theta^{jb} H_{abc} x^{c} \mathcal{A} \left( 
\tau_{1},\tau_{2} \right),
$$

\noindent
therefore yielding a correction to the noncommutative $\theta$
parameter as,

\begin{equation}
\theta^{ij} \to \theta^{ij} + {\frac{1}{3}} \theta^{ia} \theta^{jb} H_{abc}
x^{c}.
\end{equation}

\noindent 
To this result we now add the correction to the two--point function which
goes as $\mathcal{B}^{2}$,

$$
- {\frac{i }{3\pi}} G^{ia} G^{jb} H_{abc} x^{c} \mathcal{A} \left( 
\tau_{1},\tau_{2} \right).
$$

\noindent 
These two results produce the full correction to the noncommutative
parameter,

\begin{equation}  \label{correction-theta}
\theta^{ij} \to \theta^{ij} - {\frac{1}{3}} G^{ia} G^{jb} H_{abc} x^{c} + 
{\frac{1}{3}} \theta^{ia} \theta^{jb} H_{abc} x^{c}.
\end{equation}

\noindent 
Finally, there are also mixed corrections going as $\mathcal{A}\mathcal{B}
$ (and also the ``symmetric'' $\mathcal{B}\mathcal{A}$). They are,

$$
- {\frac{1 }{3\pi}} \theta^{ia} G^{jb} H_{abc} x^{c}
\mathcal{B} \left( \tau_{1},\tau_{2} \right),
$$

\noindent
plus the symmetric contribution in $i$ and $j$. These contributions yield
the correction to the effective open string metric,

\begin{equation}  \label{correction-metric}
G^{ij} \to G^{ij} - {\frac{1}{3}} \theta^{ia} G^{jb} H_{abc} x^{c} + 
{\frac{1}{3}} G^{ia} \theta^{jb} H_{abc} x^{c}.
\end{equation}

\noindent
All these results can be nicely combined with the flat spaces formulas
which connect closed string and open string parameters \cite{Schomerus,
Seiberg-Witten}, to yield new but identical formulas in this curved
background scenario.

\subsection{Open String Parameters}

Let us first recall how, in the flat case, one relates open and closed
string parameters \cite{Schomerus,Seiberg-Witten}:

\begin{eqnarray}  \label{open-parameters}
\frac{1}{G} &=& {\frac{1}{g+\omega}}\ g\ {\frac{1}{g-\omega}}\ ,  \notag \\
\theta &=& - {\frac{1}{g+\omega}}\ \omega\ {\frac{1}{g-\omega}}\ .
\end{eqnarray}

\noindent
In here, $\omega = B + F$ is constant, $G_{ij}$ is the metric
effectively seen by the open strings and $\theta^{ij}$
is the noncommutativity parameter on the brane world--volume (we never
consider noncommutativity along the time direction), $[x^{i}, x^{j}] =
i\theta^{ij}$.

Let us now consider the above formulae (\ref{open-parameters}) with
$\omega$ replaced with the curved background expression,

$$
\widetilde{\omega}_{ab} (x) = \omega_{ab} + {\frac{1}{3}} H_{abc} x^{c},
$$

\noindent 
and let us expand  (\ref{open-parameters}) to first order in
$H$. Denoting by $\widetilde{G}$ and $\widetilde{\theta}$
the ``curved'' open string parameters, one obtains:

\begin{eqnarray}
{\frac{1}{\widetilde{G}}} + \widetilde{\theta} &=& {\frac{1}{g + \omega + 
{\frac{1}{3}}Hx}} \notag \\
&\simeq& {\frac{1}{g+\omega}} - {\frac{1}{g+\omega}}\ 
{\frac{1}{3}} H x\ {\frac{1}{g+\omega}}  \notag \\
&=& \left( {\frac{1}{G}} - {\frac{1}{G}}\ {\frac{1}{3}} H x\ \theta -
\theta\ {\frac{1}{3}} H x\ {\frac{1}{G}} \right) + \left( \theta - 
{\frac{1}{G}}\ {\frac{1}{3}} H x\ {\frac{1}{G}} - \theta\ {\frac{1}{3}} 
H x\ \theta \right).  \notag
\end{eqnarray}

\noindent 
It is clear that we have just obtained the previous results
(\ref{correction-metric}) and (\ref{correction-theta}). Therefore, formulas
(\ref{open-parameters}) are still valid in the curved background situation,
but now the fields are taken to be varying fields rather than constant
fields. In other words, formulas (\ref{open-parameters}) are still valid
for weakly varying non--closed gauge invariant two--form
$\widetilde{\omega}$.

In particular, in the zero slope $\alpha^{\prime}\to 0$ limit of 
\cite{Seiberg-Witten}, the effective open string parameters are given by

\begin{equation}  \label{zero-slope}
{\frac{1}{\widetilde{G}}} = - {\frac{1}{\widetilde{\omega}}}\ g\ 
{\frac{1}{\widetilde{\omega}}}\ , \ \ \ \ \ \ \widetilde{\theta} = 
{\frac{1}{\widetilde{\omega}}}\ .
\end{equation}

\section{Tachyons and Matrix Models in Curved Backgrounds}

In this section we analyze in some detail the zero--point function, 
{\it i.e.}, the partition function, and connect the discussion
of this paper, in this simple case, to some known results in Matrix
models. 

Let us start by considering the Born--Infeld action in the presence of a
weak background field $H$. We shall be brief, since the arguments which
follow are very well known. One starts by expanding the determinant in
order to obtain

$$
S = \int \sqrt{ \det \left(\delta_{ab}+\frac{1}{3} H_{abc} x^c + 
F_{ab}\right) } \simeq \int \left(1 + \frac{1}{4} F^2 +
\frac{1}{6} H_{abc} x^c F_{ab} \right) .
$$

\noindent
We can then use the canonical correspondences $F_{ab} \to i [X^a,X^b]$ 
and $\int \to {\rm Tr}$ in order to rewrite the RHS above as

\begin{eqnarray} \label{action}
S &\simeq& {\rm Tr} \left(1 -\frac{1}{4} g_{ac}g_{bd} [X^a,X^b] 
[X^c,X^d] +
\frac{i}{3} H_{abc} X^a X^b X^c \right) \nonumber \\
&\simeq& {\rm Tr} \left(1 +
\frac{i}{3} H_{abc} X^a X^b X^c \right) + {\mathcal O}(g^2) .
\end{eqnarray}

\noindent
Note that the above action has been found by \cite{ARS-1, ARS-2} in the 
context of studies of branes in WZW models at large level $k$ ---that is, 
at small $H_{abc} \sim k^{-1/2}$--- and by Myers in \cite{Myers} in the 
context of studies of polarization of lower--dimensional branes in the 
presence of R--R background fields. In the sequel we will just look at the 
terms in the above equation which are independent of $g_{ab}$, and 
concentrate on the linear terms in $H_{abc}$. A special case of the results 
of this paper is the zero--point function, or partition function,

\begin{equation} \label{partition}
Z = P[1] =  {\rm Tr} \left(1 - \frac{2i}{3} H_{abc} X^a X^b X^c 
\right) .
\end{equation}

\noindent
At first sight, there is an incompatibility between equations 
(\ref{action}) and (\ref{partition}), since one expects that $Z \sim S$.
We have, on the other hand, a difference

\begin{equation} \label{difference}
S-Z \simeq   H_{abc} {\rm Tr} \left(X^a X^b X^c \right) .
\end{equation}

\noindent
Recall though, see \textit{e.g.} \cite{Witten-BSFT, Shatashvili-BSFT, 
Tseytlin-BSFT}, that the partition function and the action need not be 
equal and are expected to differ by a renormalization group beta 
function contribution. More precisely,

$$
S = (1+\beta\frac{\partial}{\partial g}) Z .
$$

\noindent
We will show that the difference (\ref{difference}) is nothing but this 
extra term, thus resolving the apparent contradiction.

Recall that the coefficient $-2i/3$ in equation (\ref{partition}) was fixed 
in section 4 in order to obtain conformally invariant results. This 
corresponded to an ill defined vacuum graph (which contributes to the
volume form) which is 
linearly divergent. In this paper we have chosen a {\it specific} 
regularization scheme which preserves the conformal invariance of 
the results. This scheme, however, does {\it not} correspond to the usual 
minimal subtraction scheme, as we will show in a moment, and contributes a 
finite part to the tachyon beta function, thus explaining the 
difference (\ref{difference}).

Let us be more specific. Let us consider the more general boundary 
interaction $S_B$, including the tachyon field,

$$
\int d\tau \left[ \frac{1}{2\pi }T_B\left( X\right) +iA_{a}\left( 
X\right) \dot{X}^{a}\right] .
$$

\noindent
In the above, $T_B$ is the {\it bare} tachyon field, given by
$T_B = T+\Delta T$, where $\Delta T$ are the tachyon counterterms.
Now let us consider the vacuum graph in question and let us
regularize it following, \textit{e.g.}, the prescription of 
\cite{Tseytlin-BSFT}. The graph then contributes (including the tachyon 
counterterm):

\begin{equation} \label{tachyon}
- \Delta T -\frac{i}{6}H_{abc}x^{c}\int d\tau \, \langle \zeta 
^{a}\dot{\zeta}^{b} \rangle .
\end{equation}

\noindent
Working on the disk and regularizing \cite{Tseytlin-BSFT} the result, one finds 
that:

$$
\int d\tau \, \langle \zeta ^{a}\dot{\zeta}^{b} \rangle =
2i\theta ^{ab}\frac{e^{-2\varepsilon }}{1-e^{-2\varepsilon }} 
= i\theta ^{ab}\left[ \frac{1}{\varepsilon } - 1 + {\cal O}\left( \varepsilon 
\right) \right] .
$$

\noindent
Therefore one obtains that equation (\ref{tachyon}) yields a result of

$$
-\Delta T - \frac{i}{3}H_{abc}X^a X^b X^c \left[ \frac{1}
{\varepsilon }-1\right] ,
$$

\noindent
where we have used that $\theta^{ab} = -i [X^a,X^b]$. The usual 
prescription is the one of minimal subtraction, {\it i.e.}, the counterterm 
just cancels the pole leaving a finite result of $\frac{i}{3} H_{abc} X^a X^b 
X^c$. We choose, on the other hand, a different renormalization prescription 
dictated by conformal invariance, which gives as a total contribution 
$-\frac{2i}{3} H_{abc} X^a X^b X^c$. This implies that the tachyon counterterm 
must be

$$
\Delta T= -i\left( \frac{1}{3\varepsilon }-1\right)
H_{abc} X^a X^b X^c  ,
$$

\noindent
and that the corresponding beta function is

$$
\beta _{T}=-T-H_{abc}X^a X^b X^c.
$$

\noindent
Following the methods of \cite{Witten-BSFT, Shatashvili-BSFT, 
Tseytlin-BSFT}, this contribution to the beta function implies that the 
total action, including the tachyon potential, is given by

$$
{\rm Tr} \left[ \left( 1+T+\frac{i}{3}H_{abc} X^a X^b X^c \right) 
e^{-T} \right] ,
$$

\noindent
which is extremized at $T=-\frac{i}{3}H_{abc} X^a X^b X^c$. The value
of the action at the extremum is then

$$
{\rm Tr} \left(1+ \frac{i}{3}H_{abc} X^a X^b X^c\right) ,
$$

\noindent
thus showing that the difference between the partition function and the
action is compensated by a condensation of the tachyon.

\subsection{11--Dimensional Language and the 3--form Field}

In order to be complete, one still needs to relate the previous Matrix
theory action, which is written in the 10--dimensional type IIA language,
to the full $M$--theory 11--dimensional language. Using the 11-dimensional
light--cone notation and the previous operator form of the action, we have
actually built a Matrix theory action in a weakly curved background.  One
can use the ideas from \cite{Sen, Seiberg}, and their application to curved
backgrounds \cite{Taylor-Raamsdonk-2}, to make precise the relation between
the Matrix theory and the $D$--brane Born--Infeld actions. We shall now
briefly look at these issues, with a particular attention to the
11--dimensional 3--form field.

Let us start by considering $M$--theory with a background metric $g_{IJ}$,
in a frame with a compact coordinate $X^{-}$ of size $R$, which is
light--like in the flat space limit $g_{IJ} \to \eta_{IJ}$.  This theory
can be described as the limit of a family of space--like compactified
theories \cite{Sen, Seiberg}. Define the theory $\widetilde{M}$ with
background metric $\widetilde{g}_{IJ}$ in a frame with a space--like
compact direction $X^{10}$ of size $\widetilde{R}$. The DLCQ limit of the
original theory, $M$, can be found by boosting the theory $\widetilde{M}$
in the $X^{10}$ direction with boost parameter,

$$
\gamma = {1\over \sqrt{1-\beta^{2}}} = \sqrt{1+{1\over 2} \left( {R 
\over \widetilde{R}} \right)^{2}},
$$

\noindent
and then taking the limit $\widetilde{R} \to 0$. The 3--form field
$\widetilde{A}_{IJK}$ in the theory $\widetilde{M}$ is related to that of
the original theory $M$ by the same Lorentz transformation as
above. Moreover, the theory $\widetilde{M}$ on a small space--like circle
of radius $\widetilde{R}$ is equivalent to type IIA string theory with
background form fields given by:

$$
\widetilde{A}_{IJK} d\widetilde{X}^{I} \wedge d\widetilde{X}^{J} 
\wedge d\widetilde{X}^{K} = C^{D2}_{\mu\nu\rho} dX^{\mu} \wedge 
dX^{\nu} \wedge dX^{\rho} + B_{\mu\nu} dX^{\mu} \wedge dX^{\nu} \wedge 
dX^{10}.
$$

The configurations of interest in this paper carry no $D0$ or $D2$--brane
charge. One will thus be left with a $B_{\mu\nu}$ background form field
which will give rise to the following $M$--theory background 3--form field
(here we take $\alpha \equiv \gamma (1-\beta)$):

$$
A_{+-i} = 0\ , \ \ \ A_{ijk} = 0\ , \ \ \ 
A_{+ij} = {1\over\sqrt{2}\alpha} B_{ij}\ , \ \ \ 
A_{-ij} = -{\alpha\over\sqrt{2}} B_{ij}\ , 
$$

\noindent
where one should recall that we have only space--space $B$--field turned on.

It is interesting to observe that, at the $M$--theory level, the
nonassociative Kontsevich deformation obtained is associated to the
$A_{+ij}$ and $A_{-ij}$ components of the 11--dimensional field $A_{IJK}$
(see also \cite{CHL} for the standard Moyal deformation). One is therefore
led to speculate that, from a purely $M$--theoretic point of view, one
should be able to construct deformations associated to 3--index tensor
structures, which should reduce to Kontsevich type of deformations for
configurations considered in this paper. This is an interesting venue to
explore, as it may also aid in understanding brane world--volume
deformations associated to non--zero varying R--R fields (and R--R field
strengths). One thing one can say is that 3--index tensor structures will
probably be naturally associated to 3--component products or 3--brackets in
the sense that one can write, given some (constant) tensor $C^{ijk}$,

$$
\{ f,g,h \} (x) = e^{{i \over 2} C^{ijk} \partial^{x}_{i} \partial^{y}_{j}
\partial^{z}_{k}} f(x) g(y) h (z) |_{x=y=z}.
$$

\noindent
Structures involving 3--brackets have been previously discussed in the
context of covariant Matrix theory actions \cite{Minic}. It would be
interesting to further explore these ideas.

\section{Future Perspectives}

In this paper we have shown how to use open string perturbation theory in
order to describe brane physics in weakly curved backgrounds. The method
described allows one to translate the properties of the curved background
---which traditionally show up as sigma model couplings--- into a given
deformation of the algebra of functions on the brane world--volume, which
depends in general on the specific closed string background considered. In
particular, the presence of an NS--NS field strength $H$ induces a
nonassociative deformation of the algebra of functions.

Our choice of background is of the type $R + \frac{1}{4} H^{2} = 0$, and it
would be interesting to further develop the disk perturbation theory in
order to investigate the properties of the star product deformation to
higher order. It would also be interesting to study tachyon condensation in
such a background. Given that one can compute correlators using star
product prescriptions (as thoroughly explained in this work), one can then
use standard boundary string field techniques in order to compute the
minima of the tachyon potential in this background.

Another interesting point is to further study the $\bullet$ product.
Defining gauge theory on these ``nonassociative manifolds'' is not
straightforward. Also, given the discussion about Matrix theory in curved
backgrounds, it seems clear that understanding the geometry of these
``nonassociative manifolds'', much like there is a geometrical
understanding of Kontsevich's noncommutative manifolds \cite{BFFLS,
Fedosov, JSW}, would be needed in order to fully understand Matrix theory
in any given background. More pragmatically, we were able to map functions
to matrices because we wrote everything in terms of the $\star$ product,
which is associative. A question that immediately rises is whether there is
a ``matrix'' formulation of the theory which can be written exclusively in
terms of the $\bullet$ product. Answering this question could be of great
interest for the goal of defining Matrix theory in general curved
backgrounds. This definitely requires a full understanding of the role of
homotopy associative algebras. We hope to address some of these questions
in the near future.

\vs{5}
\noindent
{\small 
{\bf Acknowledgments:}
We would like to thank P.~Fonseca, G.~Granja, J.~Mour\~ao, C.~Nappi,
J.~Nunes, C.~Nu\~nez and B.~Pioline for helpful comments and/or
discussions. LC is funded by a European Postdoctoral Institute
fellowship. RS is supported in part by funds provided by the Funda\c c\~ao
para a Ci\^encia e Tecnologia, under the grant Praxis XXI BPD-17225/98
(Portugal).
}

\eject

\appendix

\section{Dilogarithm Identities}

Let us recall that the standard dilogarithmic function is defined by the
following expression:

$$
\mathrm{Li}_2\left( x\right) =- \int_{1}^{x}\frac{\ln \left( 1-s\right) }{s} 
ds=\sum_{n=1}^{\infty }\frac{1}{n^{2}} x^n.
$$

\noindent
In particular\textrm{\ }$\mathrm{Li}_2\left( 0\right) =0$ and $\mathrm{ 
Li}_2\left( 1\right) =\zeta \left( 2\right) =\frac{1}{6}\pi ^{2}$. A
related function, more useful for our purposes, is the Rogers normalized
dilogarithm \cite{Rogers} $L\left( x\right) $ defined for $0\leq x\leq 1$ by 

$$
L\left( x\right) =\frac{6}{\pi ^{2}}\left[ \mathrm{Li}_2\left( x\right) + 
\frac{1}{2}\ln \left( x\right) \ln \left( 1-x\right) \right] .
$$

\noindent
The Rogers dilogarithm is monotonically increasing on $\left[ 0,1\right] $
and, at the end points of the interval, is given by

$$
L\left( 0\right) =0\ , \ \ \ \ \ \ \ L\left( 1\right) =1.
$$

\noindent
The function $L$ satisfies a fundamental property, which is crucial in our
computations, namely, given $x,y\in \left[ 0,1\right] $ the following holds:

\begin{equation}
L\left( x\right) +L\left( y\right) =L\left( xy\right) +L\left( \frac{x\left(
1-y\right) }{1-xy}\right) +L\left( \frac{y\left( 1-x\right) }{1-xy}\right) .
\label{app1}
\end{equation}

\noindent
A special important case of the above equation is Euler's identity,

$$
L\left( x\right) + L\left( 1-x\right) =1.
$$

\section{Computation of the Function $S\left( x\right) $}

Let us start the computation of $S(x)$ by analyzing an auxiliary function,
which we shall denote by $s\left( x\right)$, and which will be defined for 
all $x\in \mathbb{R}$. Let us consider three points $0$, $1$ and $x$ on the 
real line and let us denote, given a point $p$ in the upper--half plane, with 
$\alpha$, $\beta$ and $\gamma$ the angles formed by the lines $p$--$0$, 
$p$--$1$ and $p$--$x$ with the vertical line. A little plane geometry shows 
that:

\begin{equation}
\tan \gamma =\left( 1-x\right) \tan \alpha +x\tan \beta . \label{app2}
\end{equation}

\noindent
The function $s$ will then be given by:

$$
s\left( x\right) = \frac{4}{\pi ^{3}}\int_{\Sigma }\gamma d\alpha \wedge
d\beta .
$$

\noindent
From the geometric construction it is clear that 

$$
s\left( 1-x\right) = - s\left( x\right) .
$$

\noindent
We also recall that the upper--half plane $\Sigma $ corresponds to the
simplex $-\frac{\pi }{2}\leq \alpha \leq \beta \leq \frac{\pi }{2}$ in the 
$\alpha $--$\beta $ plane. This fact can be used to compute special values of 
$s$,

\begin{eqnarray}
s\left( +\infty \right) &=& - s\left( -\infty \right) =1\ , \nonumber \\
s\left( 1\right) &=& - s\left( 0\right) =\frac{1}{3}\ , \label{app3}
\end{eqnarray}

\noindent
and $s\left( 1/2 \right) = 0$. It is clear that, from the definition of 
$S$, one has for $x\in \left[ 0,1\right] $, 

\begin{equation}
S\left( x\right) =s\left( \frac{1}{x}\right) -s\left( x\right) +s\left( 
\frac{-x}{1-x}\right) .  \label{app4}
\end{equation}

\noindent
Consider the derivative $\frac{d}{dx}s\left( x\right) $. From (\ref{app2})
we deduce that:

$$
\frac{d}{dx}\gamma =\frac{\tan \beta -\tan \alpha }{1+\left( \left(
1-x\right) \tan \alpha +x\tan \beta \right) ^{2}}\ ,
$$

\noindent
so that,

\begin{eqnarray*}
\frac{d}{dx}s &=&\int_{\Sigma }\frac{d}{dx}\gamma d\alpha \wedge d\beta = \\
&=&\int_{-\infty <z<w<\infty }\frac{dzdw}{\left( 1+z^{2}\right) \left(
1+w^{2}\right) }\frac{w-z}{1+\left( \left( 1-x\right) z+xw\right) 
^{2}}\ ,
\end{eqnarray*}

\noindent
where we have defined $z=\tan \alpha $, $w=\tan \beta $. The above integral
can be evaluated with the result:

$$
\frac{d}{dx}s=-\frac{1}{\pi ^{2}}\left[ \frac{\ln \left( x\right) 
^{2}}{1-x} + \frac{\ln \left( 1-x\right) ^{2}}{x}\right] .
$$

\noindent
The above equation can be easily integrated by noting that 
$\frac{d}{dx} \mathrm{Li}_2\left( x\right) =-\frac{\ln x}{1-x}$. Using the 
boundary values (\ref{app3}) one obtains the following expression for 
$S$,

\begin{eqnarray*}
s &=&-\frac{1}{3}+\frac{4}{\pi ^{2}}\left[ \mathrm{Li}_2\left( x\right)
+\frac{1}{2}\ln \left( -x\right) \ln \left( 1-x\right) \right]\ ,\ \ \ \ \ \
\ \ \ \ \ \ \ \ \ \ \ \left( x<0\right) \\
s &=&\frac{2}{\pi ^{2}}\left[ -\mathrm{Li}_2\left( 1-x\right) +\mathrm{ 
Li}_2\left( x\right) \right]\ ,\ \ \ \ \ \ \ \ \ \ \ \ \ \ \ \ \ \
\ \ \ \ \ \ \ \ \ \ \ \ \ \ \ \ \ \left( 0<x<1\right) \\
s &=&\frac{1}{3}-\frac{4}{\pi ^{2}}\left[ \mathrm{Li}_2\left( 1-x\right) + 
\frac{1}{2}\ln \left( x\right) \ln \left( x-1\right) \right]\ .\ \ \ \ \ \ \
\ \ \ \ \ \ \ \ \ \ \left( x>0\right)
\end{eqnarray*}

\noindent
We finally use equation (\ref{app4}) and the fact that $\mathrm{Li}_2 
\left( 1-1/x\right) =-\mathrm{Li}_2\left( 1-x\right) -\frac{1}{2}\left( \ln
\left( x\right) \right) ^{2}$ to show that 
\begin{eqnarray*}
S\left( x\right) &=&\frac{6}{\pi ^{3}}\left( -\mathrm{Li}_2\left(
x\right) +\mathrm{Li}_2\left( 1-x\right) \right) \\
&=&-L\left( x\right) +L\left( 1-x\right) \\
&=&1-2L\left( x\right) .
\end{eqnarray*}

{\small {\ }}

\eject

\end{document}